\documentclass[twocolumn]{aastex62}
\usepackage[utf8]{inputenc}
\usepackage{units}
\usepackage{hyperref}
\usepackage{multirow}

\usepackage{CJK}
\usepackage[T1]{fontenc}

\shorttitle{Astrophysical Binary Neutron Stars}
\shortauthors{Zhu \& Ashton}

\begin{document}
\begin{CJK}{UTF8}{gbsn}

\title{Characterizing Astrophysical Binary Neutron Stars with Gravitational Waves}
\correspondingauthor{Xing-Jiang Zhu}
\email{zhuxingjiang@gmail.com}

\author[0000-0001-7049-6468]{Xing-Jiang Zhu (朱兴江)}
\affiliation{School of Physics and Astronomy, Monash University, Clayton, VIC 3800, Australia}
\affiliation{OzGrav: Australian Research Council Centre of Excellence for Gravitational Wave Discovery, Clayton, VIC 3800, Australia}

\author[0000-0001-7288-2231]{Gregory Ashton (艾格瑞)}
\affiliation{School of Physics and Astronomy, Monash University, Clayton, VIC 3800, Australia}
\affiliation{OzGrav: Australian Research Council Centre of Excellence for Gravitational Wave Discovery, Clayton, VIC 3800, Australia}

\begin{abstract}
Merging binary neutron stars are thought to be formed predominantly via isolated binary evolution.
In this standard formation scenario, the first-born neutron star goes through a recycling process and might be rapidly spinning during the final inspiral, whereas the second-born star is expected to have effectively zero spin at merger.
Based on this feature, we propose a new framework for the astrophysical characterization of binary neutron stars observed from their gravitational wave emission.
We further propose a prior for the dimensionless spins of recycled neutron stars, given by a gamma distribution with a shape parameter of 2 and a scale parameter of 0.012, extrapolated from radio pulsar observations of Galactic binary neutron stars.
Interpreting GW170817 and GW190425 in the context of the standard formation scenario and adopting the gamma-distribution prior, we find positive support (with a Bayes factor of 6, over the nonspinning hypothesis) for a spinning recycled neutron star in GW190425, whereas the spin of the recycled neutron star in GW170817 is small and consistent with our prior.
We measure the masses of the recycled (slow) neutron stars in GW170817 and GW190425 to be $1.34_{-0.09}^{+0.12}$ $\unit[(1.38_{-0.11}^{+0.11})]{M_{\odot}}$ and $1.64_{-0.11}^{+0.13}$ $\unit[(1.66_{-0.12}^{+0.12})]{ M_{\odot}}$, with $68\%$ credibility, respectively.
We discuss implications for the astrophysical origins of these two events and outline future prospects of studying binary neutron stars using our framework.
\end{abstract}


\section{Introduction}
The ground-based gravitational-wave interferometers Advanced LIGO \citep{aLIGO} and Advanced Virgo \citep{aVirgo} have discovered dozens of compact binary coalescence events \citep{GWTC1,GW190412,GW190814,gw190425} and numerous candidates\footnote{\url{https://gracedb.ligo.org/superevents/public/O3/}}.
These include binary black holes, binary neutron stars (BNSs), and possibly neutron star-black hole mergers\footnote{The recently published event, GW190814, contains either the lightest black hole or heaviest neutron star \citep{GW190814}.}, which are revolutionizing our understanding of the Universe.

Current gravitational-wave inference methods \citep{Veitch_2015, pycbcinference, Bilby19} label the two merging compact objects as primary and secondary, with corresponding masses $m_1 \geq m_2$.
However, components of BNS mergers are expected to be of comparable masses, which makes the $(m_{1},m_{2})$ parameterization inadequate.
It has long been thought that neutron star (NS) masses in BNS systems are restricted within a narrow range around $\unit[1.35]{M_{\odot}}$ \citep{Ozel12,Kiziltan13}.
Such a perception was gradually changed with a steadily increased sample of observed pulsars in Galactic BNSs in recent years.
There are now 19 known Galactic BNS systems; 12 of them have masses measured for both stars (see Table \ref{tab:BNSspin} in Appendix \ref{sec:appenA} for details).
The most asymmetric merging Galactic BNS has a mass ratio $q=0.78$ \citep{Ferdman20}, although see \citet{Andrews20} for a recent claim that $98\%$ of merging BNSs are expected to have $q>0.9$ if the Galactic BNS population is representative.
Looking at all available precise NS mass measurements, which mostly come from observations of binary pulsars  \citep[e.g.,][]{AntoniadisMSPmass16,Alsing18_NSmass}, the minimum and maximum are $\unit[1.17]{M_{\odot}}$ \citep{Martinez15} and $\unit[2.14]{M_{\odot}}$ \citep{Thankful20NSmass}, respectively.
Whereas not necessarily corresponding to the natural extremes of NS masses, they provide a reasonable lower limit at $q \approx 0.55$ for BNS mergers.
The ability to measure the mass ratio of compact binary mergers through gravitational waves depends sensitively on our prior knowledge of the spins of the merging objects.
For the first BNS merger event GW170817, \citet{GW170817properties} found a 90\% credible lower bound of $q=0.53$ if both stars are allowed to possess extreme spins.
Such a lower bound is comparable to the $q\approx 0.55$ limit arising from current pulsar mass measurements.
Below, we describe a new astrophysically motivated framework, where the two stars in BNS mergers are distinguished by type\footnote{We provide a brief overview of pulsar phenomenology and define our terminology in Appendix \ref{sec:appenA}.} rather than by mass.
This allows us to analyze NS mergers with an astrophysical prior.

It is believed that the BNS merger rate is dominated by the standard formation channel of isolated binary evolution \citep{Flannery75,DeLoore75,Smarr76,Massevitch76,Kalogera07,PostnovLRR14,Tauris17_BNS}, since the merger rate of alternative dynamical formation is orders of magnitude lower \citep{Phinney91rate,Bae14,Belczynski18,Ye20_BNSrate}.
In the standard formation scenario, the two stars of the BNS system have distinct properties.
The first-born NS is expected to undergo a recycling process where it gets spun up by accreting matter from its companion star \citep{Radhak82,Alpar82,Sriniva82,vanHeuvel17}, prominently during the case BB Roche-lobe overflow \citep{Delgado81,Tauris15}.
The end product is a recycled NS, with a spin period of order 10 to 100 ms and low spin-down rate.
The second-born NS, on the other hand, is ``normal": it spins down quickly, in $\sim \unit[10]{Myr}$ from a birth spin period of tens of ms to $\unit[\mathcal{O}(1)]{s}$.
For gravitational-wave analysis, the second born NS is effectively nonspinning during the final inspiral, and hence termed as the slow NS.
In the astrophysical parameterization, the masses (dimensionless spin) of the recycled and slow NSs are denoted as $m_r$ ($\chi_r$) and $m_s$ ($\chi_s$), respectively. We do not impose any ordering on $m_r$ and $m_s$.

Among the Galactic BNS population, the Double Pulsar (J0737$-$3039A/B) is an excellent example for our \textit{recycled} and \textit{slow} labeling scheme for BNS mergers.
It is unique, with both NSs being observed as radio pulsars \citep{Burgay03,Lyne04}: one is recycled with a spin period ($P$) of $\unit[22.7]{ms}$ and a spin-down rate ($\dot{P}$) of $\unit[1.76\times 10^{-18}]{s s^{-1}}$, and the other is slow with a spin period of $\unit[2.8]{s}$ and a spin-down rate of $8.92\times 10^{-16}$.
They are expected to merge in $\unit[86]{Myr}$.
There are another nine BNSs that will merge within a Hubble time.
All pulsars but one observed in these BNSs are recycled, with spin periods from 17 to $\unit[76.5]{ms}$, and spin-down rates from $1.6\times 10^{-19}$ to $8.6\times 10^{-18}$.
The exception is PSR J1906+0746, which is a young pulsar with a spin period of $\unit[144]{ms}$.
It is spinning down quickly ($\dot{P}=2\times 10^{-14}$) and will become a slow pulsar when merging with its companion in $\unit[300]{Myr}$ \citep{PSR1906}.

The majority of known Galactic BNSs are found in the Galactic disk and can be described in the standard formation scenario \citep[e.g.,][]{Tauris17_BNS}.
Among 10 merging BNSs in the Galaxy, PSR B2127+11C is the only exception, being found in the globular cluster M15 \citep{PSRB2127C90}.
\citet{PhinneySigurdsson91} and \citet{Prince91} suggested that the original stellar companion of PSR B2127+11C might be replaced by another NS through a dynamical encounter, which also resulted in the ejection of the BNS to the outskirts of M15.
It was estimated that the BNS merger rate from globular clusters similar to M15 can account for a significant fraction ($\sim 10-30\%$) of observed short gamma-ray bursts \citep{Grindlay06,Lee10}.
However, \citet{Ye20_BNSrate} found that dynamical interactions in globular clusters are dominated by black holes and thus make a negligible contribution to the overall BNS merger rate \citep[see also][]{Belczynski18}.
Therefore, it is generally believed that the BNS merger rate inferred from observations by the LIGO/Virgo Collaboration \citep[LVC;][]{gw170817,gw190425} is dominated by the standard formation channel \citep{Mapelli18,Neijssel19}, although see \citet{Andrews19} for speculation of a dynamical origin for several Galactic-disk BNSs with similar orbital characteristics to PSR B2127+11C.

\citet{Farrow19_BNSmass} analyzed mass measurements of Galactic BNSs in the $(m_{r},m_{s})$ framework and found modest evidence for distinct distributions of $m_r$ and $m_s$, and a bimodal distribution of $m_r$.
These features might be due to different supernova explosion mechanisms \citep{Schwab10,ThompsonSN12}, or the recycling process \citep[but probably to a much lesser extent, see, e.g.,][]{Tauris17_BNS}.
\citet{Farrow19_BNSmass} demonstrated that dozens of new BNS observations, which are achievable with future observing runs of Advanced LIGO/Virgo and ongoing/new radio pulsar surveys in $\lesssim \unit[5]{years}$, are required to draw firm conclusions.
In this work, we extrapolate from observational properties of Galactic BNS systems to derive an astrophysical prior on $\chi_r$.
This will allow us to probe distributions of $m_r$ and $m_s$ through gravitational-wave observations.

The remainder of this paper is organized as follows.
In Section \ref{sec:spin}, we derive an astrophysical prior distribution on $\chi_r$, demonstrate the robustness of the $\chi_{s}=0$ assumption, and briefly summarize our prior knowledge on spin tilt angles of recycled NSs in BNS systems.
In Section \ref{sec:2events}, we apply the new population prior to the analysis of two BNS mergers published so far -- GW170817 \citep{gw170817} and GW190425 \citep{gw190425} -- using public data releases from LVC. We also propose a couple of novel tests for the dynamical formation hypothesis.
Last, we discuss astrophysical implications from our analysis for both events and outline future prospects in Section \ref{sec:conclu}.

\section{Spins of binary neutron stars}
\label{sec:spin}

\subsection{An astrophysical prior on the dimensionless spins of the first-born recycled neutron stars}
\label{sec:mr_spin}

\begin{figure}[ht]
\begin{center}
  \includegraphics[width=0.46\textwidth]{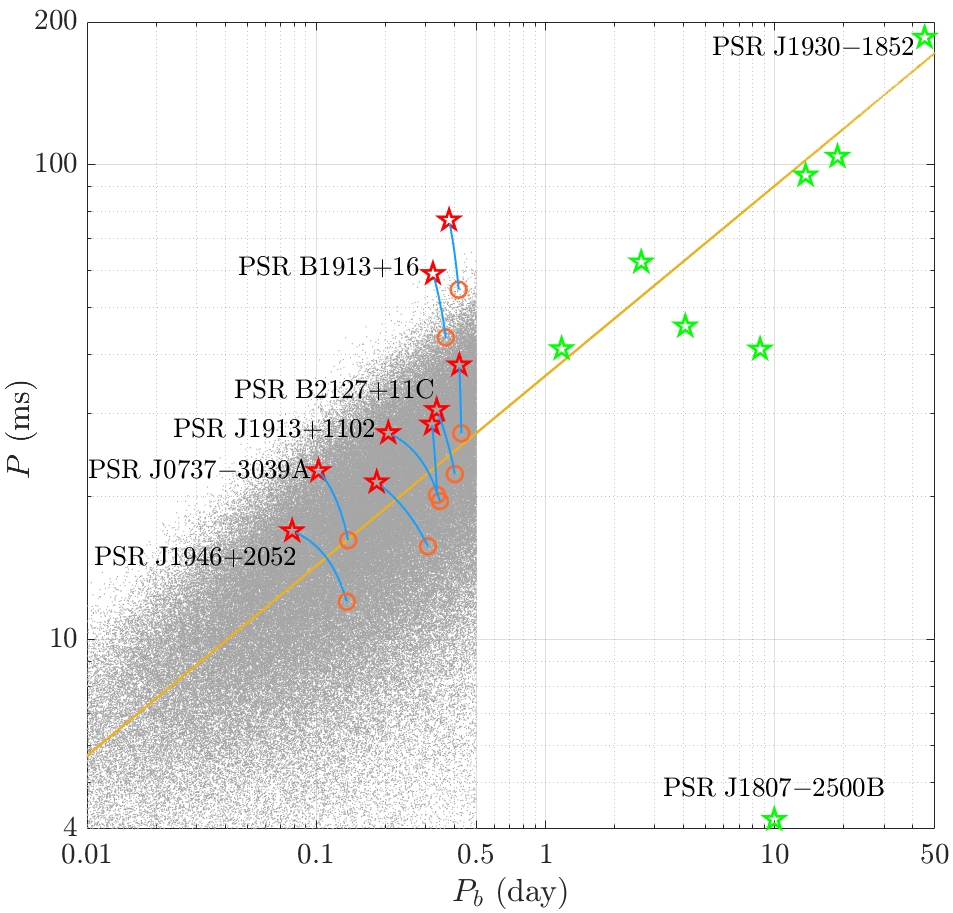}
  \caption{The $P$-$P_b$ correlation (Equation \ref{eq:P0Pd}) for BNS systems formed via isolated binary evolution. Also shown are 17 Galactic BNSs that contain an observed recycled pulsar. Red (green) stars indicate merging (non-merging) systems. Orange circles mark plausible initial states (at the birth of the second NS) for merging systems, with blue lines showing the evolutionary tracks backward in time up to half the characteristic ages ($P/2\dot{P}$) of the recycled pulsars. Next to star symbols are listed names of some pulsars mentioned in the text. Grey dots in the background represent synthetic BNS systems in this work. Two BNSs in globular clusters are shown for completeness.
  \label{fig:PPb}}
\end{center}
\end{figure}

Here, we establish an astrophysical prior on the dimensionless spin magnitudes, $\chi_r$, of recycled NSs in BNS systems, as measured in the final binary inspiral.
Following \citet{Zhu18_BNSspin}, we extrapolate from the observed properties of $10$ Galactic BNSs to a representative population of BNSs, each characterized by $m_r$, $m_s$, initial orbital period $P_b$ and eccentricity, and $P$ and $\dot{P}$ of the recycled NS at the birth of second NS.
Observationally, the spin periods of recycled pulsars in Galactic-disk BNS systems appear to be correlated with their orbital periods (see Figure \ref{fig:PPb}).
To generate our birth BNS population, we make use of the following empirical $P$-$P_{b}$ correlation \citep{Tauris15,Tauris17_BNS}:
\begin{equation}
\label{eq:P0Pd}
P=(36 \pm 14) \, {\rm{ms}}\, (P_{b}/{\rm{days}})^{0.4}\,.
\end{equation}
The above equation was originally obtained by \citet{Tauris17_BNS} in an empirical fit to 11 Galactic-field BNS systems, with a particular weight placed on the widest binary, PSR J1930$-$1852, which has a spin period of $\unit[186]{ms}$ and an orbital period of $\unit[45]{day}$ \citep{Swiggum1930}.
In Figure \ref{fig:PPb}, one can see that this correlation is supported by the up-to-date sample of observations.
Since we are primarily interested in a reasonable prior on $\chi_r$, we ignore the uncertainty on the correlation slope and hold it fixed at $0.4$.

Equation~(\ref{eq:P0Pd}) suggests that binaries born with tighter orbits tend to contain faster-spinning recycled NSs, which can be (qualitatively) attributed to a longer recycling process \citep[see][for details]{Tauris17_BNS}.
To obtain the distribution of $P$ at merger, we follow the spin-down evolution of each recycled NS in our population from the birth of the BNS to binary merger.
We adopt the standard magnetic dipole braking model \citep{Goldreich69,Ostriker69,Spitkovsky06} for spin-down evolution, and the formalism of \citet{Peters64} for gravitational-wave-driven binary orbital evolution.
The initial spin-down rate is determined by assigning a log-normal distribution of magnetic field strength that covers the range of measurements for Galactic BNSs, given an NS equation of state or radius parameter.
The magnetic field strength is assumed to be constant from binary birth to merger.
Whereas magnetic field decay might occur for young NSs \citep{Goldreich92} or for NSs undergoing the recycling process \citep{TaamHeuvel86,Romani90}, it is thought to be unlikely for recycled NSs such as millisecond pulsars \citep{BransLevin18}.
More details on the methodology and prescriptions of the BNS population can be found in \citet{Zhu18_BNSspin}.

\begin{figure}[ht]
\begin{center}
  \includegraphics[width=0.46\textwidth]{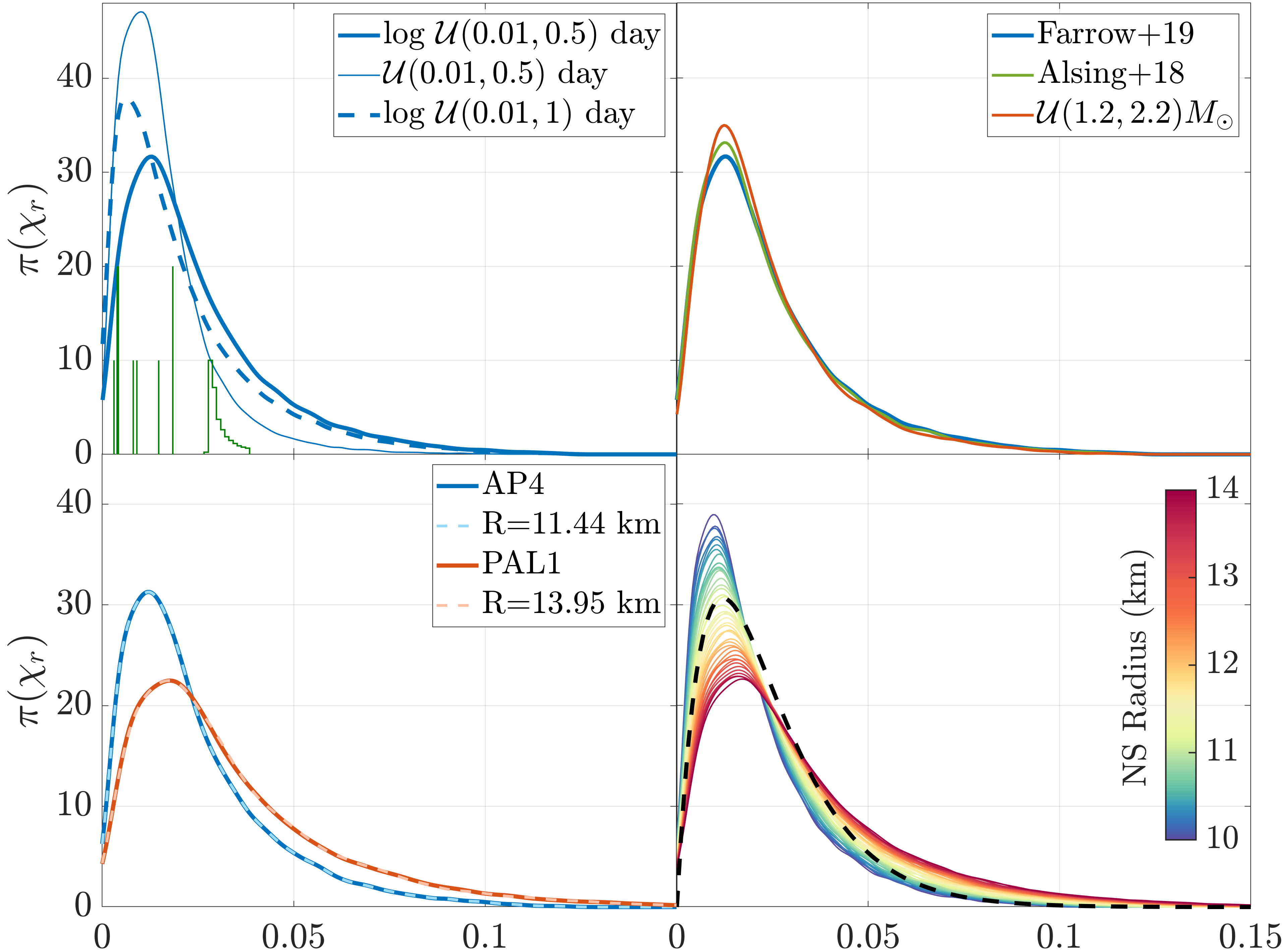}
  \caption{The probability distribution of the dimensionless spins $\pi(\chi_r)$ of recycled NSs in BNS systems.
  \textit{Top-left}: different initial binary orbital period distributions ($\mathcal{U}$ -- uniform distribution, and log $\mathcal{U}$ -- uniform distribution on a $\log_{10}$ scale), along with expected $\chi$ during merger for nine recycled pulsars in merging Galactic BNSs (green lines, scaled to a coordinate height of 10 for each NS -- a height of 20 indicates two pulsars having identical $\chi$ at merger).
  \textit{Top-right}: different mass distributions -- Farrow+19 \citep{Farrow19_BNSmass}, Alsing+18 \citep{Alsing18_NSmass}, and uniform between 1.2 and $\unit[2.2]{M_{\odot}}$.
  \textit{Bottom-left}: two NS equations of state (AP4 and PAL1) and under the assumption that all NSs have the same radius (R=11.44 km and 13.95 km). Note that there exist subtle differences between using an equation-of-state model and a single NS radius parameter; solid and dashed lines precisely overlap as a result of curve smoothing.
  \textit{Bottom-right}: distribution of $\chi_r$ for a range of NS radii from 10 to 14 km (colored lines), and the gamma distribution with a shape parameter of 2 and a scale parameter of 0.012 (black dashed line). We note that our spin distribution model shown here is qualitatively similar to more sophisticated population synthesis models \citep[e.g.,][]{Debatri20}.
  \label{fig:mr_spin}}
\end{center}
\end{figure}

The definition of $\chi\propto I/(m^{2}P)$ means that there are three ingredients in the distribution of $\chi$: 1) the distribution of spin periods ($P$), 2) the distribution of masses ($m$), and 3) the equation of state, which, together with $m$, determines the moment of inertia ($I$).
To compute the moment of inertia, we make use of the empirical relation between $I/(mR^2)$ and $m/R$ (with $R$ being NS radius) found in \citet{LattimerSchutz05} for a range of realistic NS equations of state (see their equation 12).
As a side note, we find that a more natural choice of prior would be on $P$, so that the mass distribution and NS equation of state can be simultaneously inferred with a population of events \citep[e.g.,][]{Wysocki20}.
Nevertheless, in what follows we describe a model for the distribution of $\chi_r$ that takes into account uncertainties in the distributions on $P$ and $m$ and the equation of state.
This will provide insights into how much information about the NS mass distribution and equation of state can be extracted from spin measurements \textit{alone} through hierarchical inference.

In Figure~\ref{fig:mr_spin}, we show the effect of different initial orbital period ($P_b$) distributions (top-left panel), mass distributions (top-right panel), and the NS equation of state (bottom two panels) on the distribution of $\chi_r$.
In our fiducial model, we assume a log-uniform distribution of $P_b$ between 0.01 and 0.5 days.
In comparison, the orbital periods of 10 merging Galactic BNSs range from 0.078 to 0.421 days.
By applying the $P$-$P_{b}$ correlation down to an orbital period of $\unit[0.01]{days}$, the shortest initial spin period is about $\unit[4]{ms}$.
Overall, around $10\%$ of recycled NSs have $P< \unit[10]{ms}$ at the birth of the second NS in our population.
This makes our model conservative since theoretical modeling of standard BNS formation suggests that the first-born NS is usually only moderately recycled; \citet{Tauris15} found a shortest spin period of $\unit[11]{ms}$ assuming an accretion efficiency of three times the Eddington limit during the case-BB Roche-lobe overflow.

Based on measurements of Galactic BNSs, we adopt a two-Gaussian distribution of $m_r$, peaking at $1.34$ and $\unit[1.47]{M_{\odot}}$, with width of $0.02$ and $\unit[0.15]{M_{\odot}}$ and weight of $0.68$ and $0.32$, respectively \citep{Farrow19_BNSmass};
the distribution of $m_s$ is uniform between $1.1$ and $\unit[1.5]{M_{\odot}}$.
The distribution of $m_s$ is only needed for calculation of the merger time and we find it has no impact on our results.
Our fiducial choice of equation of state is AP4 \citep{Lattimer01EOS}.

In the top-left panel of Figure~\ref{fig:mr_spin}, we also show the result for two alternative distributions of $P_b$: a) uniform between 0.01 and 0.5 days (thin solid line), and b) log-uniform between 0.01 and 1 days (thick dashed line).
In comparison to our fiducial model (thick solid line), both a) and b) result in a smaller fraction of fast-spinning NSs, because recycled NSs have smaller initial spins in long-$P_b$ binaries and the spin-down time is longer (meaning smaller residual spins at merger).
Even though a) and b) may appear to provide a better fit to Galactic BNSs (shown as green lines), we choose our fiducial model because it allows a relatively larger fraction of fast spins.
This is a conservative choice from the perspective of building a population prior.
It is reasonable to believe \textit{a priori} that the actual fraction of fast-spinning recycled NSs is greater than that of the \textit{observed} Galactic BNS population because of at least three selection effects.
The first two arise from the short orbital periods associated with fast recycled NS spins (i.e., the $P$-$P_{b}$ correlation).
Pulsars in these tight binaries are more difficult to detect in radio pulsar surveys because of severe Doppler smearing of pulse signals and their short merger times.
A third one is the selection effect against short spin periods due to dispersion smearing of pulse signals by the interstellar medium, favoring the intermediate range from tens to hundreds of ms \citep[see, e.g., figure 11 in][]{Lazarus15arecibo}.

In the top-right panel of Figure~\ref{fig:mr_spin}, we show the spin distribution for three mass models.
In addition to the model of \citet{Farrow19_BNSmass}, we also consider: 1) the two-Gaussian model of \citet{Alsing18_NSmass} fitted to mass measurements of NSs in all binaries, peaking at $1.34$ and $\unit[1.80]{M_{\odot}}$, with width of $0.07$ and $\unit[0.21]{M_{\odot}}$ and weight of $0.65$ and $0.35$, respectively; and 2) a uniform distribution between 1.2 and 2.2 $M_{\odot}$.
It is apparent that the distribution of spin magnitude is insensitive to the mass model.
In the bottom-left panel, we show the distribution of $\chi_r$ for two equations of state -- AP4 and PAL1, as investigated in \citet{Zhu18_BNSspin}, and assuming that all NSs have the same radius (11.44 km and 13.95 km, respectively).
It can be seen that the effect of equation of state is completely captured by the radius parameter.

In the bottom-right panel of Figure~\ref{fig:mr_spin}, we show the spin distribution (depicted in different colors) for a range of NS radii from 10 to 14 km -- a plausible range allowed by current gravitational-wave and pulsar observations \citep{Landry20NSradius}.
As an arbitrary fit to the group of colored curves, the black dashed curve is given by the gamma distribution with a shape parameter of 2 and a scale parameter of 0.012.
We propose this gamma distribution to be used as an astrophysical prior on $\chi_r$ in gravitational-wave data analysis. Once dozens or more BNS events are detected, we may be able to update the prior to reveal information encoded in the spin distribution as shown in Figure \ref{fig:mr_spin} \citep[see also][]{Zhu18_BNSspin}.

\subsection{Can the second-born neutron star be spinning at a measurable rate during the final binary inspiral?}
\label{sec:ms_spin}

The spin period of pulsar B in the Double Pulsar system is $\unit[3]{s}$, corresponding to $\chi_{s}\approx 10^{-4}$.
PSR J1906+0746 is the only other young pulsar in merging Galactic BNS systems.
Its spin period of $\unit[144]{ms}$ translates to $\chi_{s}\approx 0.003$.
However, it is spinning down quickly, at $2\times 10^{-14}$.
By the time of binary merger (in $\unit[300]{Myr}$), its dimensionless spin is expected to be well below $10^{-3}$, which is effectively zero for gravitational-wave observations\footnote{In comparison, the 90\% credible upper bound on component spins of GW170817 under the low-spin ($\chi<0.05$) prior is $0.04$ \citep{GW170817properties}. We discuss the measurement of NS spins with gravitational waves in detail in Appendix \ref{sec:appenB}.}.
To examine under what conditions can the second-born NS be spinning fast at merger, we conduct the following investigation.

\begin{figure}[ht]
\begin{center}
  \includegraphics[width=0.46\textwidth]{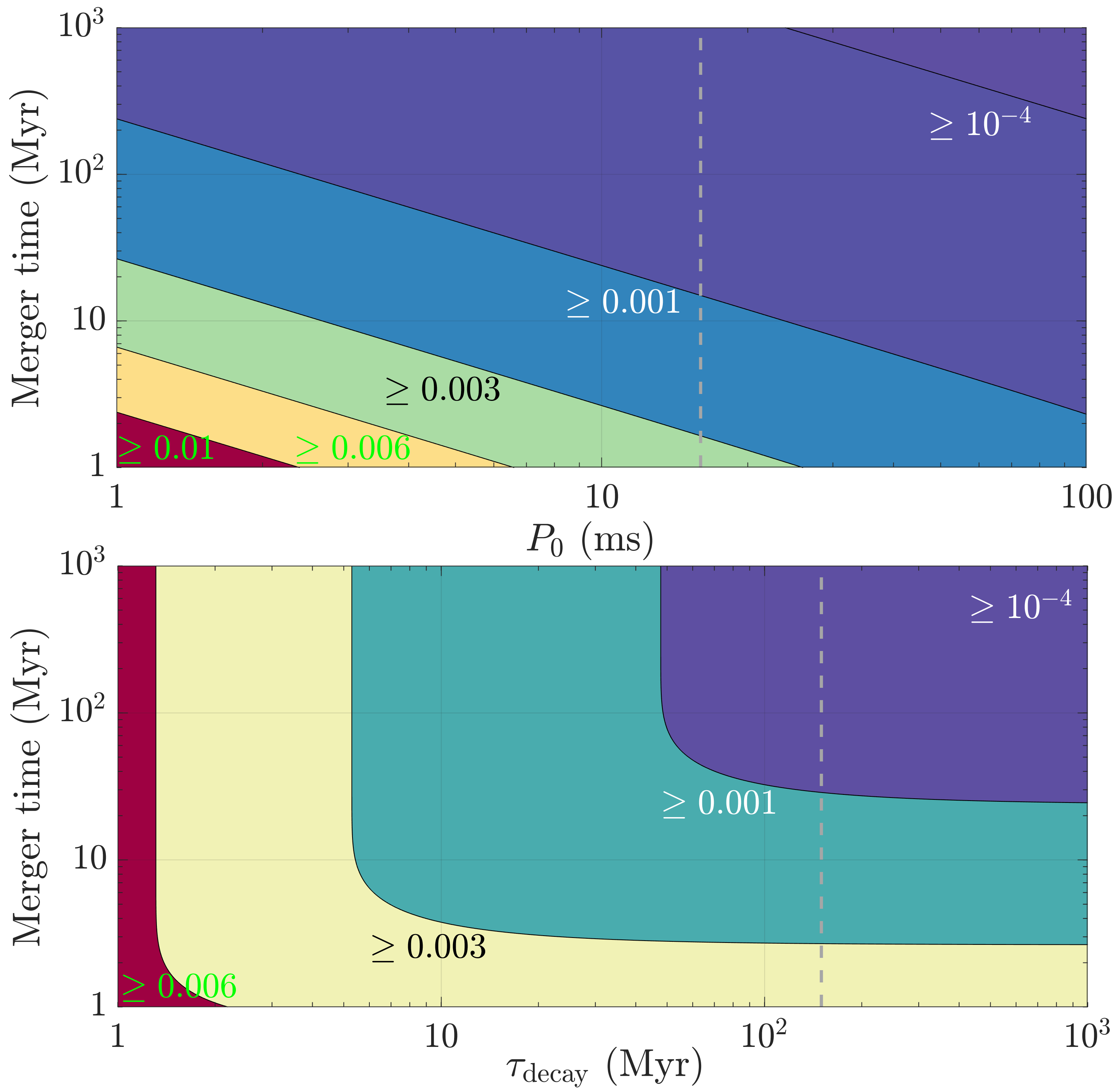}
  \caption{Lines of constant dimensionless spin ($\chi_s$) for the second-born NS in the standard BNS formation scenario: $P_0$ -- the birth spin period, $\tau_{\text{decay}}$ -- the magnetic field decay timescale. The initial spin-down rate is fixed at $2\times 10^{-14}$ in both panels. In the \textit{upper} panel, the magnetic field strength is assumed to remain constant from the NS birth to binary merger, and the vertical line marks the shortest spin period, $\unit[16]{ms}$, of known young pulsars \citep{Marshall0537}. In the \textit{lower} panel, $P_0$ is fixed at $\unit[10]{ms}$, and the vertical line indicates the a typical decay timescale of $\unit[150]{Myr}$ found in \citet{BransLevin18}.
  \label{fig:ms_spin}}
\end{center}
\end{figure}

The second-born NS in a BNS system inevitably spins down after its birth, due to the loss
of rotational energy by powering magnetically driven plasma winds \citep{Goldreich69,Contop99,Spitkovsky06}.
Its final spin period depends on the birth spin period $P_0$ and spin-down rate, and the spin-down time, which equals the binary merger time.
As we show in the previous subsection, the dimensionless spin $\chi$ is largely determined by the spin period, since the spread in NS masses and radii can change $\chi$ by no more than $\approx 50\%$.
We follow the spin-down evolution from birth to binary merger of a hypothetical NS. We assume a radius of $R=\unit[12]{km}$, and set mass $m=\unit[1.29]{M_{\odot}}$ and initial spin-down rate $\dot{P}=2\times 10^{-14}$ to those of PSR J1906+0746.
We consider two cases where the NS magnetic field strength either remains constant in this evolution or decays exponentially with a typical timescale $\tau_{\text{decay}}$ \citep[see Appendix A in][for details]{Zhu18_BNSspin}.

Figure \ref{fig:ms_spin} shows lines of constant $\chi_s$ for a range of initial spin periods, merger times, and magnetic field decay timescales.
In the upper panel, where the magnetic field is assumed not to be decaying, one can see that two conditions need to be met for $\chi_{s}\gtrsim 0.003$ -- $P_0$ in the range of a few ms and merger times shorter than $\unit[10]{Myr}$.
The vertical dashed line in this panel marks the spin period ($\unit[16]{ms}$) of PSR J0537$-$6910 \citep{Marshall0537}, the fastest-spinning young pulsar in the ATNF Pulsar Catalogue\footnote{Refer to the \href{https://www.atnf.csiro.au/research/pulsar/psrcat/}{Catalogue version 1.63}, among 50 pulsars with $P_{0}<\unit[100]{ms}$ and $\dot{P}>10^{-16}$.} \citep{ATNFpsr05}.
In the lower panel, where $P_0$ is fixed at $\unit[10]{ms}$, it becomes evident that $\tau_{\text{decay}} \lesssim \unit[10]{Myr}$ is required for $\chi_{s}\gtrsim 0.003$.
However, that is unlikely given current understanding of NS magnetic field evolution.
For example, the vertical line in the lower panel marks the typical decay timescale of $\unit[150]{Myr}$ found in \citet{BransLevin18} for young and hot NSs.

In summary, Figure \ref{fig:ms_spin} shows that the parameter space required for the second-born NS to be spinning at measurable rates during the final BNS inspiral is rather limited.
There are three additional lines of argument for $\chi_{s}=0$.
First, the spin-down rate (of PSR J1906+0746) assumed in Figure \ref{fig:ms_spin} is low for $P_{0}<\unit[100]{ms}$.
Pulsars spinning faster than PSR J1906+0746's $\unit[144]{ms}$ period generally have higher spin-down rates (e.g., $\dot{P}=5\times 10^{-14}$ for PSR J0537$-$6910).
Second, by associating young pulsars with supernova remnants and thus obtaining independent age estimates, \citet{Popov12NSspin} found that the NS birth spin periods are likely to be between 10 and hundreds of ms.
Third, even though NSs might be born with spin periods of $\mathcal{O}(\text{few})$ ms, they are expected to spin down quickly (in $\lesssim 10$ years) to $\gtrsim \unit[10]{ms}$ \citep{Lorimer93NSspin,LaiD01NSspin,Vink08}.
Therefore, we conclude that the second-born NS is effectively nonspinning for gravitational-wave observations, unless the merger time is $\ll \unit[1]{Myr}$, which we do not consider in this study.

\subsection{The spin tilt angles of recycled neutron stars}
\label{sec:spintilt}

While six parameters are required to describe the spin vectors of two NSs in a binary merger, spin effects are primarily measurable in gravitational waves through two parameters \citep[e.g.,][and references therein]{GW170817properties}: 1) the effective spin parameter $\chi_{\text{eff}}$, which is the mass-weighted combination of spins along the orbital angular momentum vector $\mathbf{L}$; 2) the effective spin-precession parameter $\chi_p$, which quantifies the spin components perpendicular to $\mathbf{L}$.
By setting $\chi_{s}=0$ and letting $\theta$ be the spin tilt angle of the recycled NS with respect to $\mathbf{L}$, it becomes evident that $\chi_{\text{eff}}= m_{r} \chi_{r} \cos\theta /(m_{r}+m_{s})$, and $\chi_{p}=\chi_{r}\sin \theta$ (if $m_{r}\geq m_{s}$) or $\chi_{p}=\chi_{r}\sin \theta (m_{r}/m_{s})(3m_{s}+4m_{r})/(4m_{s}+3m_{r})$ (if $m_{r}< m_{s}$).
Apart from $\chi_r$, the key parameter is the spin tilt angle $\theta$, for which we briefly summarize our prior knowledge below.

From a binary evolution perspective, the spin axis of a recycled NS gets aligned with orbital angular momentum during the recycling process \citep{Hills83,Bhatta91}, therefore the spin tilt angle is generally related to the supernova kick imparted on the second-born NS directed out of the orbital plane \citep{Tauris17_BNS}.
\citet{Bailes88} considered a range of BNS progenitors for PSR B1913+16 and a Rayleigh distribution with a scale parameter of $\unit[150]{km\, s^{-1}}$ for the kick velocity, and showed that $\theta \lesssim 60^{\circ}$ with $\gtrsim 90\%$ confidence.
The spin tilt angle of the pulsar can be constrained by measuring the pulse profile variations induced by spin precession.
For PSR B1913+16, $\theta$ was measured to be $18\pm 6^{\circ}$ by \citet{Kramer191316}.
In the last decade, $\theta$ was constrained for three additional recycled pulsars in Galactic BNSs: $\theta <3.2^{\circ}$ for PSR J0737$-$3039A \citep{Ferdman13DPSR}, $\theta = 27\pm 3^{\circ}$ for PSR B1534+12 \citep{FonsecaB1534}, $\theta <34^{\circ}$ for PSR J1756$-$2251 \citep{FerdmanPSR1756}.

Therefore, it is reasonable to believe that large spin tilt angles are unlikely \textit{a priori} for recycled NSs in standard BNS systems.
In the following section, we consider both aligned spins and isotropic spin orientation when we reanalyze GW170817 and GW190425 in our astrophysical framework.
With the relaxed isotropic spin prior, we are able to compare the posteriors on $\theta$ against prior knowledge mentioned here.
Unless otherwise specified, results presented hereafter assume isotropic spin orientation for the recycled NS.

\section{Reanalysis of GW170817 and GW190425}
\label{sec:2events}

Adopting our astrophysical parameterization of the masses and spins ($m_{r}$,$m_{s}$,$\chi_{r}$,$\chi_{s}$) and the spin priors derived in Section \ref{sec:spin}, we reanalyze the two BNS merger events detected by LIGO/Virgo: GW170817 and GW190425.
We use the \texttt{IMRPhenomPv2\_NRTidal} waveform model, which includes an effective description of spin-precession and tidal effects
\citep{Hannam:2013oca, Khan:2016, Dietrich2017, Dietrich:2018uni}.
In our prior, the detector-frame binary chirp mass is uniform in $[1.18,\, 1.21]$ $([1.48,1.495])\,\text{M}_{\odot}$ for GW170817 (GW190425) and the mass ratio $m_{s}/m_{r}$ is uniform in $[0.125,\, 8]$, with detector-frame component masses restricted in $\unit[[1.0,\, \unit[4.3]]]{{M_{\odot}}}$.
Note that the definition of mass ratio in our prior is necessary since we impose no ordering between $m_r$ and $m_s$. Elsewhere in the paper, e.g., when comparing with the LVC results, we adopt the $q\leq 1$ convention.
We find that our posteriors have no support near our prior limits.
Priors for other source parameters are identical to those used in the LVC discovery papers \citep{gw170817,gw190425}; for GW170817, we fix the sky location to its host galaxy NGC 4993 \citep{gw170817multi}.

\begin{figure}[ht]
\begin{center}
  \includegraphics[width=0.46\textwidth]{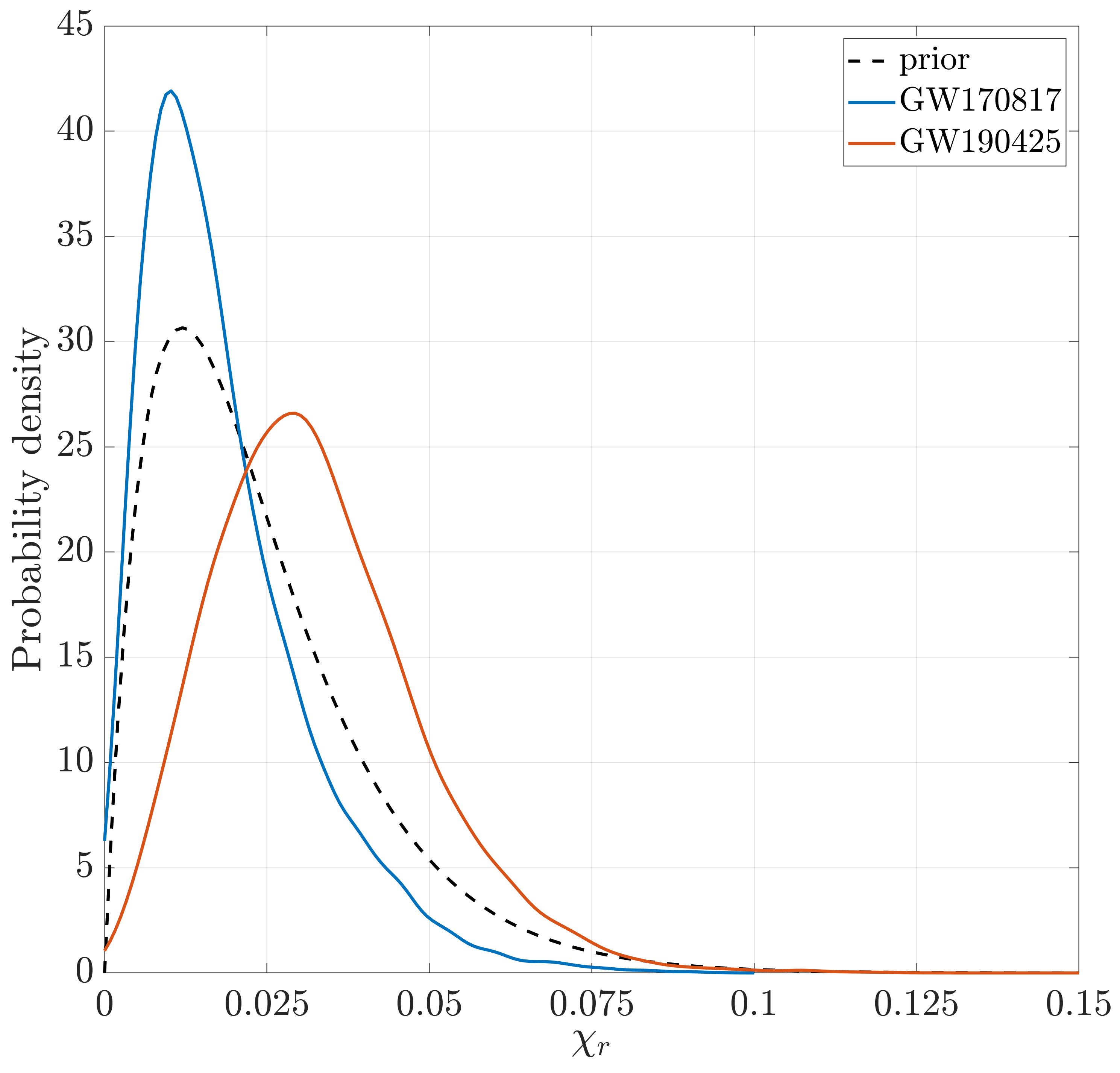}
  \caption{Posterior distributions of the dimensionless spins ($\chi_r$) of recycled NSs in GW170817 and GW190425, derived under the astrophysical prior (black dashed line) and assuming aligned spins. \label{fig:GWchi}}
\end{center}
\end{figure}

\begin{figure}[ht]
\begin{center}
  \includegraphics[width=0.46\textwidth]{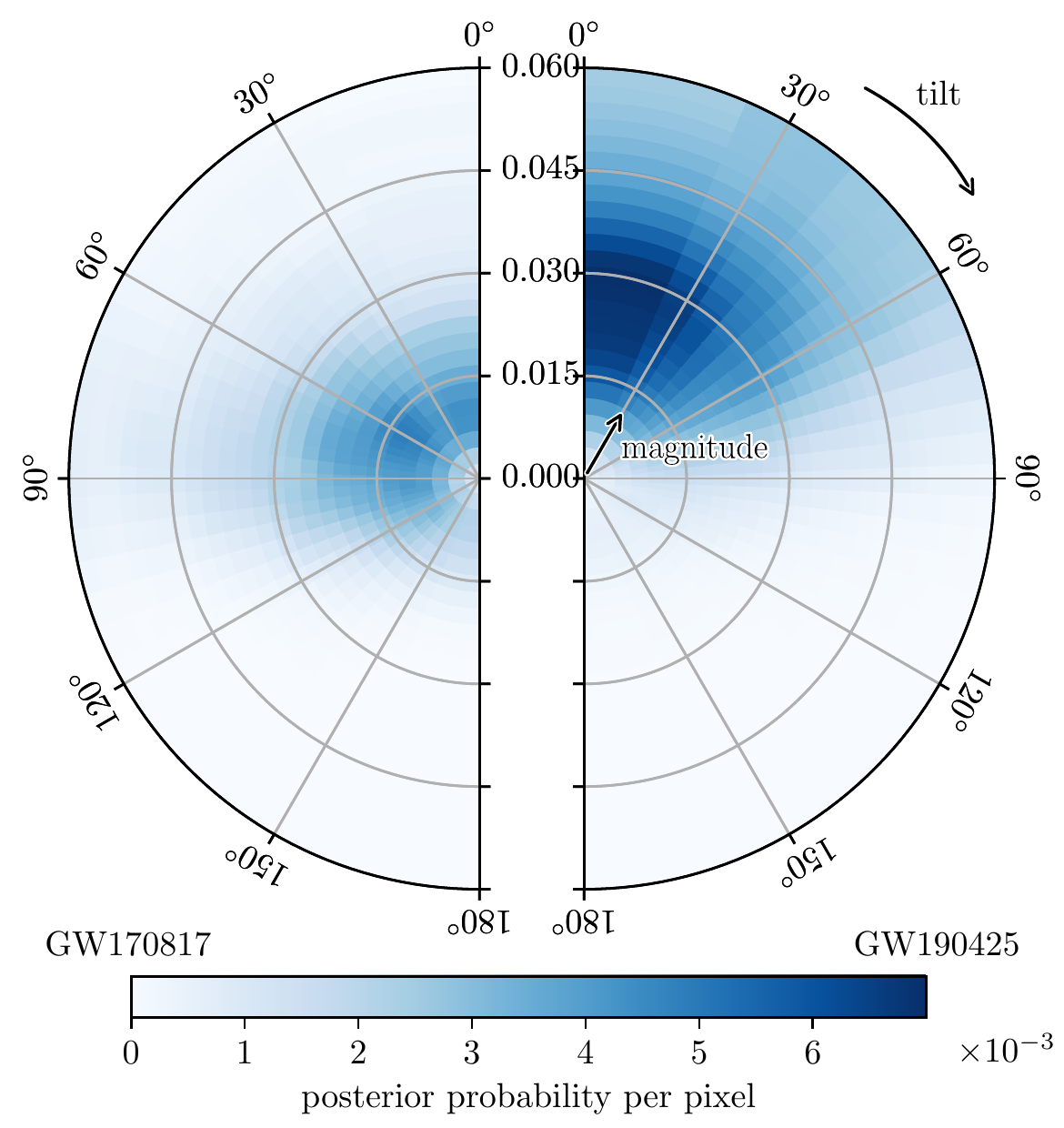}
  \caption{Posterior probability densities of $\chi_r$ in GW170817 (left) and GW190425 (right), with respect to the orbital angular momentum $\mathbf{L}$. This is shown for the astrophysical spin prior using the \texttt{IMRPhenomPv2\_NRTidal} waveform at a reference frequency of $\unit[20]{Hz}$. A tilt angle of $0^{\circ}$ indicates alignment with $\mathbf{L}$. \label{fig:chi_tilt}}
\end{center}
\end{figure}

Assuming aligned spins, we obtain a Bayes factor of 6 (0.8) between the spinning and nonspinning hypotheses for GW190425 (GW170817).
This implies positive support for NS spins in GW190425, and that there is no evidence for or against spins in GW170817.
Figure \ref{fig:GWchi} shows the posterior distributions of $\chi_r$, along with the prior.
In accordance with Bayes factor results, the posterior of GW190425 moderately shifts toward larger spins, peaking at $\chi_{r}=0.03$ in comparison to $0.012$ for the prior; for GW170817 the posterior is similar to the prior, only shifting slightly toward lower spins.

Assuming a uniform prior for $\cos \theta$ between $-1$ and $1$ (i.e., random spin orientation), Figure \ref{fig:chi_tilt} shows the posterior probability densities of $\chi_r$ with respect to the orbital angular momentum $\mathbf{L}$ for GW170817 and GW190425, on the left and right half-disks, respectively.
We find the spin in GW170817 is only weakly constrained, disfavouring anti-alignment with $\mathbf{L}$.
The weak constraint on $\theta$ is unsurprising for GW170817 because of small spins. Given its high signal-to-noise ratio ($32.4$), this adds further weight to our finding that the recycled NS in GW170817 is only mildly spinning.
On the other hand, the spin tilt angle can be constrained to be $\lesssim 60^{\circ}$ for GW190425, with strongest support below $30^{\circ}$.
This agrees well with astrophysical expectations presented in section \ref{sec:spintilt}.

\begin{figure}[ht]
\begin{center}
  \includegraphics[width=0.46\textwidth]{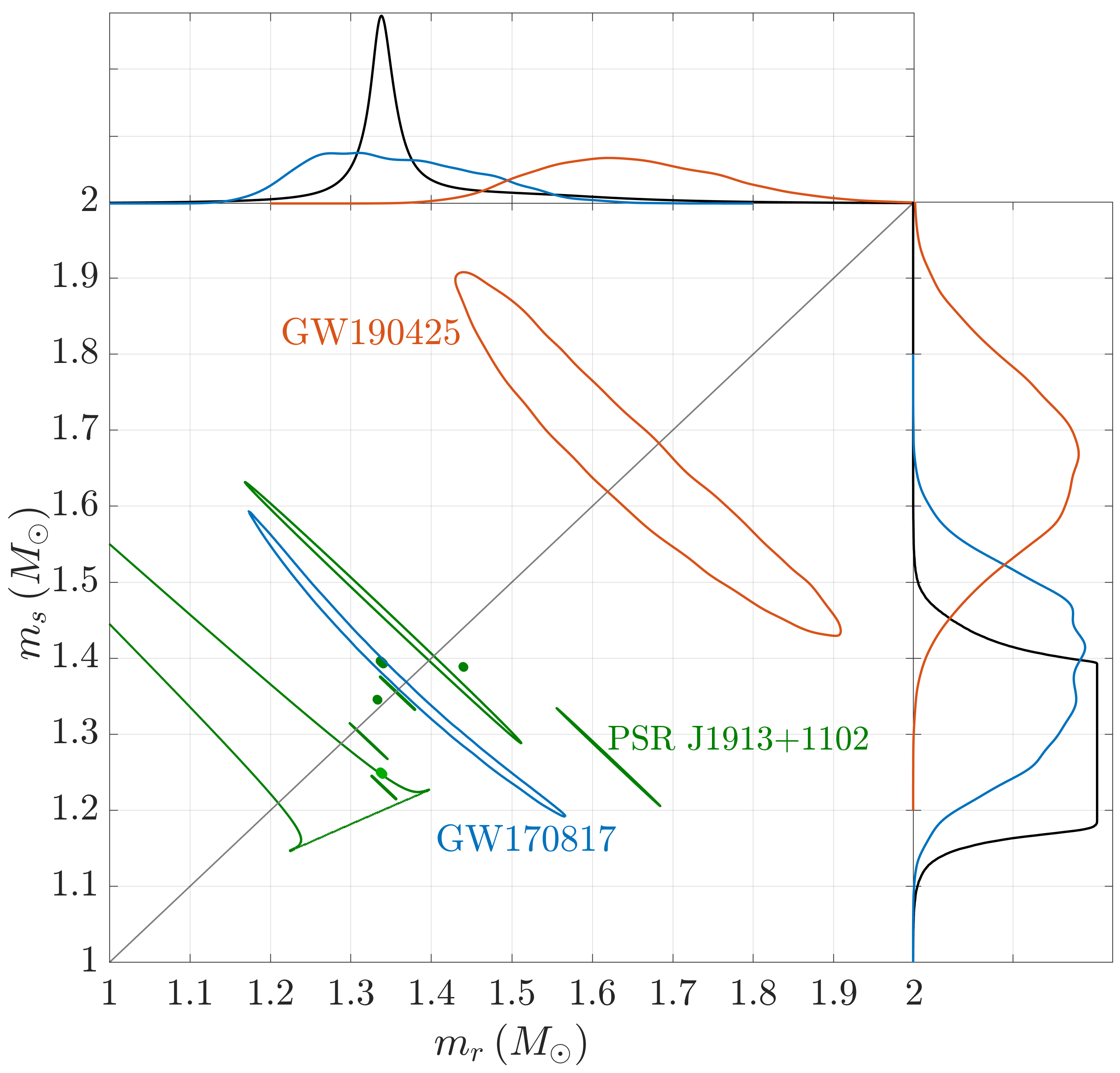}
  \caption{Posterior distributions (shown as $90\%$ credible regions) of NS masses ($m_r$ and $m_s$) of GW170817 and GW190425, along with those for 10 merging Galactic BNSs (green lines). Due to high measurement precision, four Galactic BNSs are indicated as dots in the figure. One-dimensional marginalized distributions are given in the small panels on the top and right; black curves illustrate the posterior predictive distributions based on the Galactic BNS population \citep{Farrow19_BNSmass}, where $m_r$ and $m_s$ follow a two-Gaussian distribution and a uniform one, respectively.\label{fig:mrms}}
\end{center}
\end{figure}

\begin{figure*}[ht]
\begin{minipage}[t]{0.5\textwidth}
\includegraphics[width=0.9\linewidth]{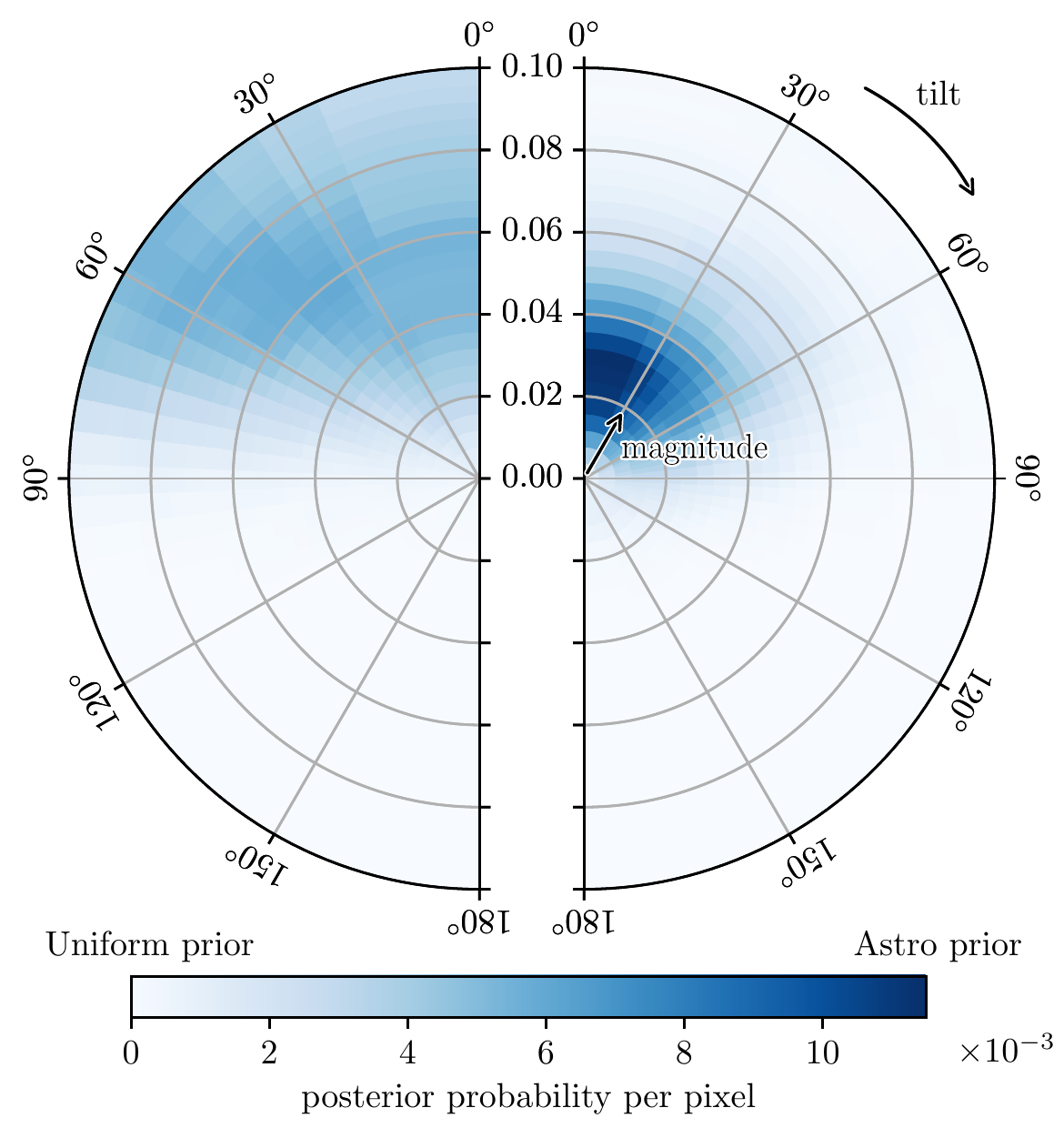}
\end{minipage}
\begin{minipage}[t]{0.5\textwidth}
\includegraphics[width=0.9\linewidth]{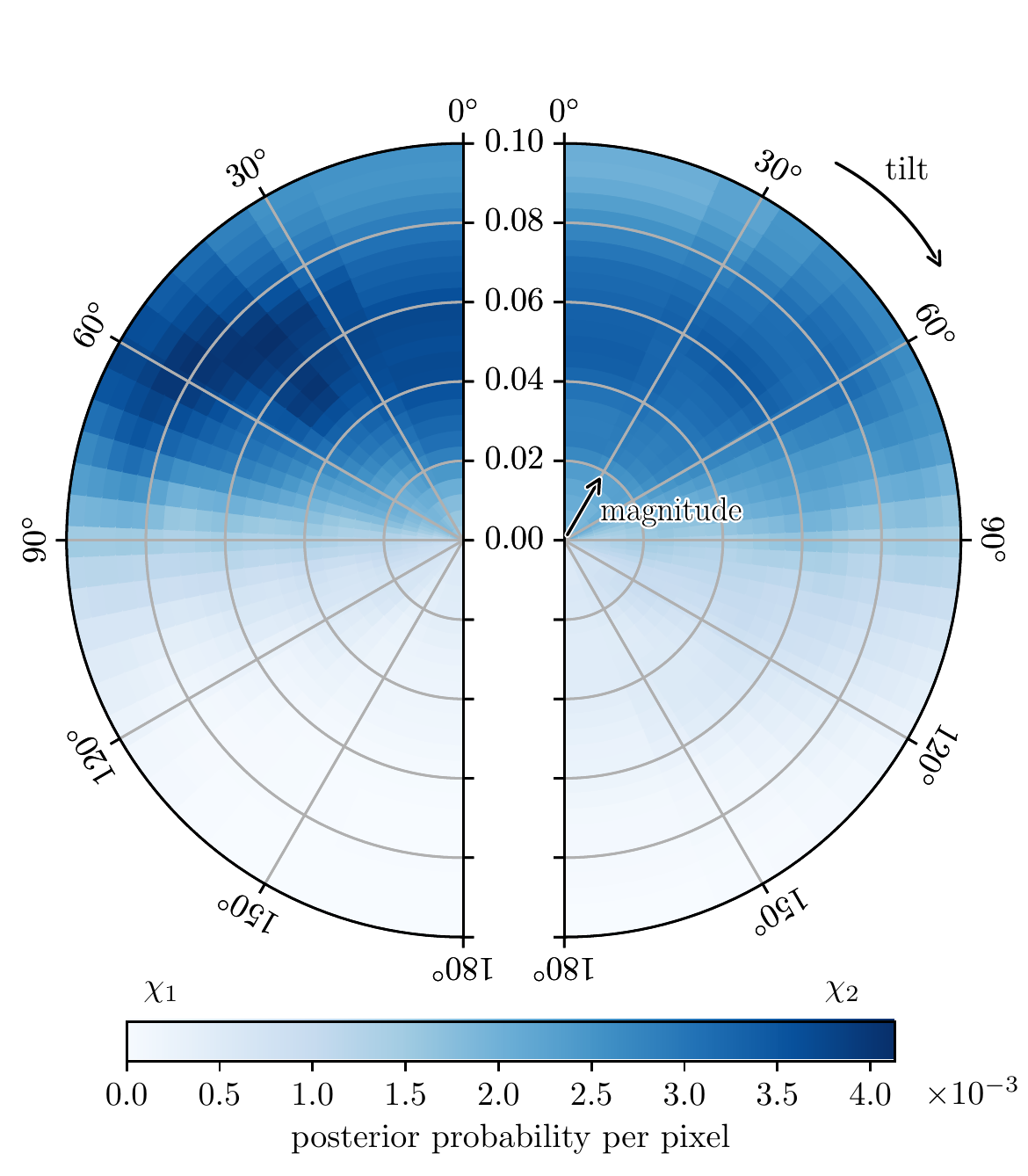}
\end{minipage}
\caption{Posterior probability densities of NS spins with respect to the orbital angular momentum for GW190425. The \textit{left} panel is shown for the recycled NS ($\chi_r$), using two different priors, uniform between 0 and 0.1, and the astrophysical prior derived in section \ref{sec:mr_spin}; note that the posterior for the astrophysical prior was already shown in Fig.~\ref{fig:chi_tilt}, but with a different magnitude range. The \textit{right} panel is shown for component spins $\chi_1$ and $\chi_2$, of the primary and secondary NSs, respectively.}
\label{fig:2spindisk}
\end{figure*}

Figure \ref{fig:mrms} shows the joint posterior distributions of $(m_{r},\ m_{s})$ for GW170817 and GW190425, along with 10 merging Galactic BNSs \citep{Farrow19_BNSmass}.
We find that both GW170817 and GW190425 are likely to be equal-mass binary mergers.
The masses of the recycled (slow) NSs are measured to be $1.34_{-0.09}^{+0.12}$ $\unit[(1.38_{-0.11}^{+0.11})]{M_{\odot}}$ and $1.64_{-0.11}^{+0.13}$ $\unit[(1.66_{-0.12}^{+0.12})]{M_{\odot}}$, with $68\%$ credibility, for GW170817 and GW190425, respectively.
Reordering $(m_{r},\ m_{s})$ into $(m_{1},\ m_{2})$, we obtain slightly improved constraints on the mass ratio (defined as $q\leq 1$).
For GW170817 (GW190425), our 90\% credible lower bound on $q$ is 0.79 (0.80), in comparison to 0.73 (0.78) by LVC\footnote{We obtained the LVC lower bounds directly from the published posterior samples \citep{GW190425_samples}. The lower bound for GW190425 was rounded to 0.8 in \citet{gw190425}.} assuming the low-spin prior ($\chi <0.05$).
We provide a comprehensive comparison with LVC results in Appendix \ref{sec:appenB}.

In the one-dimensional marginal distribution plots of Figure \ref{fig:mrms}, we show as black lines the posterior predictive distributions of $m_{r}$ and $m_{s}$ derived in \citet{Farrow19_BNSmass} for Galactic BNS systems.
We find the measurements of $m_{r}$ and $m_{s}$ for GW170817 are fully consistent with the Galactic BNS population.
The measured $m_r$ in GW190425 is broadly consistent (at the $18\%$ confidence level) with the posterior predictive distribution.
It falls under the secondary peak (at $\unit[1.47]{M_{\odot}}$ with a width of $\unit[0.15]{M_{\odot}}$) of the Galactic model, and is similar to $\unit[1.62\pm 0.03]{M_{\odot}}$ of PSR J1913+1102 \citep{Ferdman20} -- the most massive recycled pulsar in Galactic BNS systems.
However, the measurement of $m_s$ of GW190425 is inconsistent with the Galactic population ($< \unit[1.5]{M_{\odot}}$) at the $95\%$ confidence level.

\subsection{Test of the dynamical formation hypothesis}
\label{sec:test190425spin}

Our priors on NS spins derived in Section \ref{sec:spin} are valid only for the standard BNS formation channel.
While current estimates of merger rate suggest that nearly all BNS mergers should be formed via the standard channel, one detected event outside the astrophysical prior proposed in this work would indicate a new origin.
For example, in our prior, $\chi \gtrsim 0.05$ is relatively rare ($8\%$), which arises from the theoretical expectation that the first-born NS is usually only moderately recycled.
However, a dynamically formed BNS could contain a fully recycled millisecond NS ($\chi \sim 0.1-0.2$).
If both NSs were found to exhibit measurable spins, it would be a ``smoking-gun'' evidence for a dynamically formed BNS containing two recycled NSs\footnote{If formed through three-body interactions, the captured NS could also be an isolated slow NS. Therefore, the single-spinning feature is not unique to the standard formation channel.}, because that is extremely unlikely in standard BNS formation (see section \ref{sec:ms_spin}).

Since our analysis finds no support for measurable spins in GW170817, we only consider alternative possibilities for GW190425.
For this purpose, we adopt a uniform prior between 0 and 0.1 for $\chi$, where the upper end corresponds to a spin period of 3.6 (6.0) ms for an NS radius of 10 (14) km, for $m=\unit[1.65]{M_{\odot}}$. Note that PSR 1807$-$2500B, in the globular cluster NGC 6544, has a spin period of $\unit[4.19]{ms}$ \citep{Lynch12}. It is the fastest-spinning recycled pulsar in Galactic BNS systems, but the binary ($P_{b}=\unit[10]{days}$) is not expected to merge within the age of the Universe.

We find a Bayes factor of 2 between the double-spinning and single-spinning hypotheses for GW190425 (under the same uniform prior for $\chi$), and a Bayes factor of 1.3 between the uniform prior and astrophysical prior for the single-spinning hypothesis.
These small Bayes factors imply that the data are insufficient to disentangle these possibilities.
It is also worth mentioning that the Bayes factor needs to be weighted by prior odds, which are small for the dynamical formation hypothesis.
Figure \ref{fig:2spindisk} shows the posterior probability densities for the single-spinning (left) and double-spinning (right) configurations.
In the left panel, the left and right half-disks correspond to the uniform prior and astrophysical prior, respectively.
Note that in all cases except the one using the astrophysical prior, each pixel in the plot has equal prior probability.
From this test, we conclude that for GW190425: 1) anti-aligned spins (a tilt angle of $\sim 180^{\circ}$) are strongly disfavoured; 2) assuming only one NS is spinning, its spin tilt angle is $\lesssim 60^{\circ}$.

\begin{figure}[ht]
\begin{center}
  \includegraphics[width=0.46\textwidth]{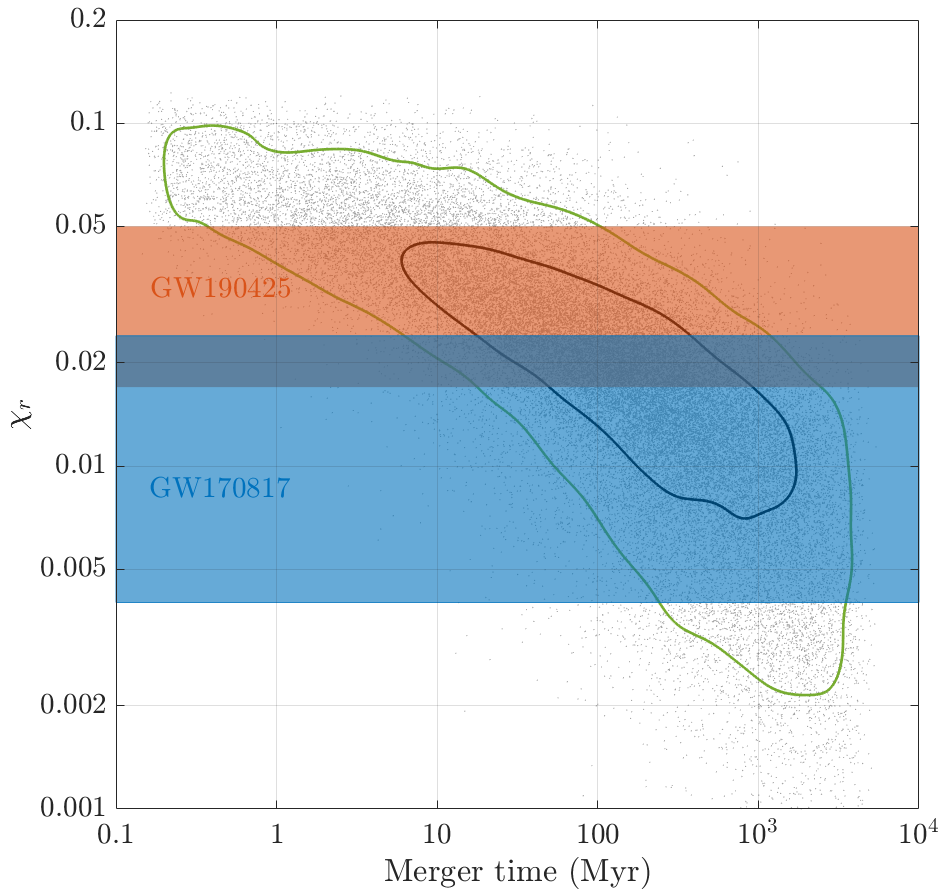}
  \caption{The correlation between BNS merger times and the dimensionless spins ($\chi_r$) of recycled NSs measured during the final binary inspiral. The two horizontal bands mark the 68\% credible intervals of the posteriors shown in Figure \ref{fig:GWchi} for GW170817 and GW190425. Black and green curves enclose 50\% and 90\% of our population prior  in the fiducial model described in section \ref{sec:mr_spin}, respectively. \label{fig:chi_Tmerge}}
\end{center}
\end{figure}

\section{Implications and future prospects}
\label{sec:conclu}

It is generally believed that the BNS merger rate is dominated by the standard isolated binary evolution formation channel.
We demonstrate that a generic feature for standard BNS formation is that the second-born NS is effectively nonspinning during the final merger.
The first-born NS, on the other hand, can get spun up during a recycling process and is likely to retain significant spin at merger.
This motivates us to propose a new \textit{recycled} and \textit{slow} labeling scheme, which also solves the dilemma of the primary-secondary parameterzation in measuring BNS mergers.

We further propose a population prior on the dimensionless spins ($\chi_r$) of recycled NSs, by extrapolating pulsar spin measurements of 10 Galactic BNS systems to a wide range of initial orbital configurations.
Such an extrapolation is made possible by the empirical correlation between the initial BNS orbital period ($P_b$) and the spin period ($P$) of the recycled NS, given by equation (\ref{eq:P0Pd}) and illustrated in Figure \ref{fig:PPb}.
This relation directly results in a correlation between BNS merger times and the final spins of recycled NSs, as shown in Figure \ref{fig:chi_Tmerge}.
It can be understood as follows.
Shorter $P_b$ implies a faster-spinning recycled NS, which also has a shorter spin-down time as determined by the merger time.
There is potentially one extra factor (not included in our population modeling) that can strengthen such a correlation -- a recycling spin-up process could reduce the NS spin-down rate.
By approximating the equilibrium spin period of a magnetized NS in the accretion process with the Kepler orbital period at the Alfv$\acute{\text{e}}$n radius \citep{GhoshLamb79}, and assuming that the NS loses its rotational energy through dipolar electromagnetic radiation, there exists a spin-up relationship $\dot{P}\sim P^{4/3}$ \citep[e.g.,][]{ArzoumanCordes99}.
Observationally, fully recycled pulsars have lower spin-down rates than recycled pulsars in Galactic BNS systems.
See Figure \ref{fig:PPdot} in Appendix \ref{sec:appenA} for illustrations.

After accounting for uncertainties in NS mass distribution and equation of state, as well as selection effects in radio pulsar surveys, we find that a representative distribution of $\chi_r$ can be approximated by a gamma distribution with a shape parameter of 2 and a scale parameter of 0.012.
Adopting the recycled-slow parameterization and the population prior on $\chi_r$, we reanalyze public LVC data for GW170817 and GW190425.
We find no support of measurable spins in GW170817, and modest evidence of spin for GW190425 with a Bayes factor of 6.
We discuss the implications of our analysis for GW170817 (section \ref{sec:discuss17}) and GW190425 (section \ref{sec:discuss25}).
Note in Figure \ref{fig:chi_Tmerge} that significant uncertainties exist for both the spin-merger time correlation and for our spin measurements.
Therefore, we use the plot to make some qualitative (instead of quantitative) statements on the astrophysical origins of those two events.
After that, we outline future prospects of studying BNS mergers using our framework.

\subsection{Implications for GW170817}
\label{sec:discuss17}
We show that the recycled NS in GW170817 is only mildly or even slowly spinning.
The small spin implies a relatively long merger time as can be seen in Figure \ref{fig:chi_Tmerge}.
Note that the sharp turnoff of merger time at $\unit[\sim 6]{Gyr}$ is due to a cut at $P_b=\unit[0.5]{days}$ applied in our population modeling, which was chosen to derive a conservative prior on $\chi_r$ (see section \ref{sec:mr_spin} for details).
With such a caveat in mind, our result is consistent with the finding of \citet{Blanchard170817tmerge}, who estimated a merger time between $6.8$ and $\unit[13.6]{Gyr}$ (90\% confidence) from the star formation history of its host galaxy NGC 4993.

We find that the two stars of GW170817 have comparable masses, around $\unit[1.35]{M_{\odot}}$.
The mass ratio is constrained to be above $0.79$ with 90\% credibility.
Recently, \citet{Ferdman20} proposed that around $2-30\%$ of BNS mergers are asymmetric, similar to the Galactic BNS system containing PSR J1913+1102 ($q=0.78 \pm 0.03$), and that GW170817 might be one such merger with large mass asymmetry.
Similar suggestions were made earlier, mostly to explain the ejecta associated with the kilonova accompanying GW170817 at optical and near-infrared wavelengths \citep[e.g.,][]{GaoHe17,Pankow170817}.
Specifically, \citet{Pankow170817} argued that the low mass ratio, $q\sim 0.65$, as required to explain kilonova ejecta via a dynamical origin \citep{Dietrich17BNSejecta,LV170817ejecta}, implies a tension between GW170817 and the Galactic BNS population.
However, \citet{Metzger18Magnetar} showed that the dynamical hypothesis fails to simultaneously explain the quantity, velocity and composition of the ejecta responsible for the luminous optical (blue) kilonova emission.
For alternative ejecta sources such as magnetar winds \citep{Metzger18Magnetar} or accretion disk outflow \citep{Siegel18}, no stringent constraint can be placed on the mass ratio of GW170817 from kilonova observations.

In short, we conclude that GW170817 is a canonical BNS merger with vanilla masses ($\approx \unit[1.35]{M_{\odot}}$) and contains a mildly spinning NS.

\subsection{Implications for GW190425}
\label{sec:discuss25}
The large total mass $\unit[3.3_{-0.1}^{+0.1}]{M_{\odot}}$ of GW190425 is inconsistent with the \textit{observed} Galactic BNS population.
\citet{gw190425} performed a simple Gaussian fit to measurements of 10 merging Galactic BNS systems and computed the inconsistency to be at the 5$\sigma$ level.
We found that when including uncertainties of model parameters for the mass distributions of Galactic BNSs, e.g., comparing to the posterior predictive distribution shown in Figure 5 of \citet{Farrow19_BNSmass}, the inconsistency is somewhat alleviated, to the 3$\sigma$ level.

Our analysis of GW190425 adds several pieces of information to that of \cite{gw190425}.
First, modest evidence of a relatively fast-spinning recycled NS points to a merger time of $\lesssim \unit[100]{Myr}$ (Figure \ref{fig:chi_Tmerge}).
Second, the constraint on the spin tilt angle $\lesssim 60^{\circ}$ is consistent with the standard formation scenario.
Third, our measurements of $(m_{r},\ m_{s})$, both around $\unit[1.65]{M_{\odot}}$, might provide additional clues to investigation of this event within the standard formation scenario.
The remaining question is, why has no such \textit{massive} BNS been observed before?
Population synthesis models are capable of reproducing the high mass of GW190425 \citep[e.g.,][]{Kruckow20,Mandelpopsyn20}, but it is unclear how the above question is tackled.
We discuss two specific proposals below.

\citet{GW190425origin} proposed that a viable formation pathway for GW190425 is the unstable case BB mass transfer process \citep{DewiPols03,Ivanova03}, which can make BNS systems born in ultratight orbits ($P_{b}\lesssim \unit[1]{hr}$). These fast-merging binaries are nearly invisible in radio pulsar surveys due to selection effects mentioned in section \ref{sec:mr_spin}.
The large mass is a result of a survival effect, because an immediate BNS progenitor (an NS-helium star binary) in a wide orbit could have produced a massive BNS but it would get disrupted during the final supernova explosion (assuming the supernova kick is proportional to the resultant NS mass).

\citet{Safarzad20} argued that the fast-merging channel is somewhat disfavoured, because it should contribute $\lesssim 10\%$ of the overall BNS merger rate, whereas the inferred merger rates based on GW170817 and GW190425 are comparable.
They invoked a NS mass-magnetic field correlation, namely, massive NSs are born with either too weak or too strong magnetic fields and thus are undetectable as pulsars.
The authors acknowledged the observational evidence of massive pulsars in binaries with white-dwarf companions \citep[e.g.,][]{Thankful20NSmass} and suggested that those systems may provide clues in understanding the origin of GW190425.

Given these unknowns, we have searched for evidence of dynamical formation for GW190425, namely, two fast-spinning NSs or only one is spinning but with a large spin outside our prior.
Unfortunately, the Bayes factors are not informative enough to tell these possibilities apart, mostly because of the relatively low signal-to-noise ratio (12.9).
Nevertheless, such a test will prove useful in revealing the formation pathways of future BNS events.

It is worth mentioning that with $10+2$ observations, we are likely to be in the regime of small-number statistics.
When comparing component masses of GW190425 to those of the Galactic population, we find the tension mostly lies in the slow-NS mass, with a reduced level of inconsistency (2$\sigma$).
Additionally, one has to keep in mind that high-mass BNS mergers are favored in gravitational waves because the detectable volume scales as $\mathcal{M}^{5/2}$ (with $\mathcal{M}$ being binary chirp mass), but they are selected against in radio because the binary lifetime scales as $\mathcal{M}^{-5/3}$.

In short, our analysis shows that GW190425 is consistent with being formed in a tight orbit (with a merger time $\lesssim \unit[100]{Myr}$), including a fast-spinning recycled NS ($\chi \sim 0.03$, or $P\sim \unit[15]{ms}$) with a spin tilt angle $\lesssim 30^{\circ}$.
Future gravitational-wave observations or radio pulsar surveys may soon unravel its ``mass mystery''.

\subsection{Future prospects}
Gravitational-wave observations are the major driving force in refining our understanding of the formation of coalescing compact binaries.
With planned upgrades of Advanced LIGO/Virgo detectors and the operation of an enlarged global network, we are likely to see weekly detections of BNS mergers within the next five years \citep{LVKprospects}.
This will usher in a new golden era in studying the formation and evolution of BNS systems.
In parallel, the pace of discovery of relativistic binary pulsars has been readily accelerating in the last few years, doubling the sample size of known Galactic BNS systems.
The radio BNS population is likely to grow significantly further, particularly thanks to new pulsar discovery and timing machines such as MeerKAT \citep{MeerKAT20} and FAST \citep{ZhangLei19fast,Cameron20fast}.

We propose a new framework for the analysis of gravitational-wave data from BNS mergers, based on the generic single-spinning feature found in the standard BNS formation scenario.
This framework naturally links gravitational-wave observations to radio pulsar observations and theoretical population syntheses.

We also develop a population prior for the dimensionless spins $\chi_{r}$ of the spinning recycled NS, using the spin period-orbital period ($P$-$P_b$) correlation and with simple assumptions on binary orbital characteristics.
There are several directions worthy of further investigations.

First, we demonstrate there exists a correlation between BNS merger times and the dimensionless spins of recycled NSs at the final inspiral.
Such a correlation can be further refined with a detailed modeling of the spin-up process and using more realistic distributions of initial binary orbital periods and eccentricities.
Second, it would be interesting to examine where the $P$-$P_b$ correlation ceases to be applicable.
In our model, we extrapolate this correlation from the shortest observed $P_b$ of $\unit[0.078]{day}$ to $\unit[0.01]{day}$ (for the sake of conservativeness), which results in a long tail beyond $\chi_{r}\approx 0.05$.
A more informed upper end for $\chi_{r}$ would help to identify atypical NSs in BNS mergers.
Third, we demonstrate that the distribution of $\chi_r$ is insensitive to NS mass distribution, but uniformly affected by the NS equation of state through the NS radius parameter.
Such insights could guide future work that aims to constrain the NS spin/mass distributions and the equation of state simultaneously via hierarchical inference.

The astrophysical framework proposed in this work will enable future gravitational-wave observations to determine whether or not the first and second-born NSs follow different mass distributions.
This will in turn have significant implications for supernova explosion mechanisms involved in BNS formation.
Once dozens of BNS mergers are detected, such a prior will also enable the measurement of the typical spin tilt angle of the recycled NS, which encodes information about the magnitudes of supernova kicks applied to the second-born NS.
A population of $\mathcal{O}(>100)$ detected events will update the prior and reveal information about the NS equation of state, the distribution of initial BNS orbital periods, and potentially magnetic field decay (if it exists).

Finally, we outline a couple of novel tests based on NS spins to probe the formation pathways of BNS mergers.
We find that a detected BNS event with measurable spins in both stars would be a smoking-gun evidence for dynamical formation.
On the flip side, a dynamically formed BNS is also likely to contain a slow NS, which could originally be an isolated NS and later be captured via dynamical interactions.
We find no support for or against either of the above two scenarios for GW190425 due to its relatively low signal-to-noise ratio.
If indeed the dynamical formation channel is not as ineffective as currently thought in producing BNS mergers, which could be confirmed via radio pulsar observations in the meantime, the NS spin tests within our framework may prove instrumental in future gravitational-wave observations.

\acknowledgments
We thank the anonymous referee for helpful comments on the manuscript.
We also thank Michael Zevin, He Gao, and Ilya Mandel for useful comments/discussions.
X.-J.Z. is supported by ARC CE170100004, G.A. is supported by ARC DP180103155.
This research has made use of data, software and/or web tools obtained from the Gravitational Wave Open Science Center (\url{https://www.gw-openscience.org}), a service of LIGO Laboratory, the LIGO Scientific Collaboration, and the Virgo Collaboration.
This work made use of the OzSTAR national facility at Swinburne University of Technology. The OzSTAR program receives funding in part from the Astronomy National Collaborative Research Infrastructure Strategy (NCRIS) allocation provided by the Australian Government.
We use the \texttt{parallel-Bilby} Bayesian inference software package \citep{Bilby19, pbilby2020} with the \texttt{Dynesty} nested sampling algorithm \citep{Dynesty}.

\appendix

\section{Pulsar phenomenology and Galactic binary neutron stars}
\label{sec:appenA}

\begin{figure}[ht]
\begin{center}
  \includegraphics[width=0.5\textwidth]{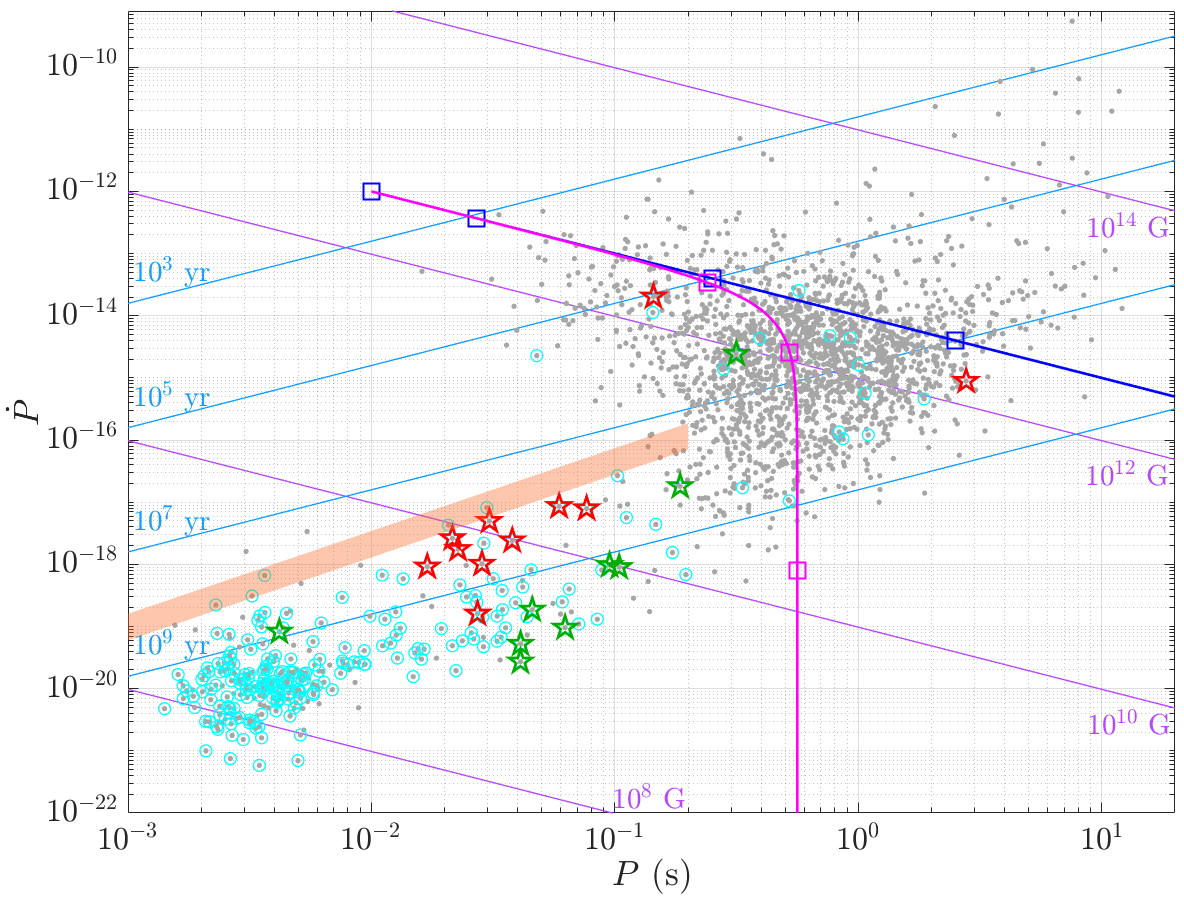}
  \caption{The pulsar $P$-$\dot{P}$ diagram based on the \href{https://www.atnf.csiro.au/research/pulsar/psrcat/}{ATNF Pulsar Catalogue Version 1.63}. Pulsars are plotted as grey dots, with cyan circles indicating binary systems. Red (green) stars mark pulsars in merging (non-merging) Galactic BNS systems. Also shown are lines of constant characteristic age $P/(2\dot{P})$ (light blue), and constant surface magnetic field strength $3.2\times 10^{19} ({\rm{Gauss}})(P\dot{P})^{1/2}$ (purple lines). As an illustration, the blue and pink line marks the evolution of a hypothetical NS, with $P=\unit[10]{ms}$ and $\dot{P}=10^{-12}$ at birth, assuming no magnetic field decay and an exponential field decay scenario with a typical timescale of $\unit[1]{Myr}$, respectively. Blue squares indicate $\{0, 10^3, 10^5, 10^{7}\}\,\text{yr}$ after birth. Pink squares indicate $\{0.1, 1, 5\}\, \text{Myr}$ after birth. The orange shaded band represents the spin-up relation $\dot{P}=(1.1\pm 0.5) \times 10^{-15}P^{4/3}$ \citep{GhoshLamb79,ArzoumanCordes99}, which approximately indicates the equilibrium spin periods of the spin-up accretion process. Note the general trend that faster-spinning recycled pulsars also have lower spin-down rates.
  \label{fig:PPdot}}
\end{center}
\end{figure}

Here we briefly summarize pulsar phenomenology relevant to BNS systems and explain some terminology used in this work through the classic $P$-$\dot{P}$ diagram (Figure \ref{fig:PPdot}).
We refer interested readers to, e.g., the book by \citet{pulsarHandbook} for an in-depth review.

Generally, NSs are believed to be born somewhere in the top-left corner of the $P$-$\dot{P}$ diagram.
Adopting the standard magnetic dipole braking model and assuming no magnetic field decay, aN NS evolves along a line of constant magnetic field strength toward the bottom-right (parallel to purple lines in Figure \ref{fig:PPdot}).
For aN NS born with a spin period of $\mathcal{O}\unit[(10)]{ms}$, it takes $\sim \unit[10]{Myr}$ to spin down to a period of $\mathcal{O}\unit[(1)]{s}$.

Pulsars inside the major cloud around $P\sim \unit[1]{s}$ and $\dot{P}\sim 10^{-15}$ in Figure \ref{fig:PPdot} are usually referred to as \textit{normal} or \textit{young} pulsars.
They are mostly isolated pulsars.
The secondary cloud, with $P$ between 1 and $\sim \unit[10]{ms}$ and $\dot{P}\sim 10^{-20}$, contains what are usually called millisecond pulsars.
Pulsars in Galactic BNSs (shown as star symbols in Figure \ref{fig:PPdot}) are mostly located in between the two clouds, except for three young pulsars and one millisecond pulsar.
Most pulsars with $P<\unit[200]{ms}$ and $\dot{P}<10^{-16}$ are found to be in binary systems, which are collectively referred to as recycled pulsars.
Depending on their spin periods, we adopt the following terminology in this paper: mildly recycled ($P\gtrsim \unit[80]{ms}$), moderately recycled ($\unit[10]{ms} \lesssim P \lesssim  \unit[80]{ms}$), and fully recycled ($P \lesssim  \unit[10]{ms}$).

\begin{table}
 \begin{tabular}{lccccccccc}
 \hline
  \hline
  Pulsar Name & $P$ (ms) & $\chi$ & $\dot{P}\,(10^{-18})$ & $P_b$ (day) & $e_0$ & $T_{\rm{m}}$ (Gyr) & $m_r$ ($M_{\odot}$) & $m_s$ ($M_{\odot}$) & Reference\\
  \hline
  J1946+2052 & 16.96  & 0.031 & 0.92 & 0.078 & 0.064 & 0.046 & \multicolumn{2}{c}{$m_{\text{tot}}=2.50 M_{\odot}$}  & \cite{Stovall18BNS}\\
  J1757$-$1854 & 21.50 & 0.023 & 2.63 & 0.184 & 0.606 & 0.076 & 1.338 & 1.395 & \cite{Cameron17BNS} \\
 J0737$-$3039A  & 22.70 & 0.022 & 1.76 & 0.102 & 0.088 & 0.086 & 1.338 & -- & \cite{Kramer06Sci}\\
 J0737$-$3039B & 2773 & $10^{-4}$ & 892 & 0.102 & 0.088 & 0.086 & -- & 1.249 & \cite{Kramer06Sci}\\
 J1913+1102 & 27.29 & 0.016 & 0.16 & 0.206 & 0.090 & 0.470 &  1.62 & 1.27 & \cite{Ferdman20} \\
  J1756$-$2251  & 28.46 & 0.017 & 1.02 & 0.320 & 0.181 & 1.656 & 1.341 & 1.230 & \cite{FerdmanPSR1756}\\
  B2127+11C  & 30.53 & 0.016 & 4.99 & 0.335 & 0.681 & 0.217 & 1.358 & 1.354 & \cite{Jacoby06}\\
  B1534+12  & 37.90 & 0.013 & 2.42 & 0.421 & 0.274 & 2.734 & 1.333 & 1.346 & \cite{FonsecaB1534}\\
 B1913+16  & 59.03 & 0.008 & 8.63 & 0.323 & 0.617 & 0.301 & 1.440 & 1.389 & \cite{Weisberg10}\\
 J0509+3801 & 76.54 & 0.006 & 7.93 & 0.380 & 0.586 & 0.574 & 1.34 & 1.46 & \cite{Lynch0509}\\
 J1906+0746  & 144.1 & 0.003 & 20268 & 0.166 & 0.085 & 0.308  & 1.322 & 1.291 & \cite{PSR1906}\\
    \hline
 J1807$-$2500B  & 4.19 & 0.117 & 0.08 & 9.957 & 0.747 & $10^{3}$ & 1.366 & 1.206 & \cite{Lynch12}\\
 J1518+4904  & 40.93 & 0.012 & 0.03 & 8.634 & 0.249 & $10^{4}$ & \multicolumn{2}{c}{$m_{\text{tot}}=2.72 M_{\odot}$} & \cite{Janssen08}\\
   J1829+2456  & 41.01 & 0.013 & 0.05 & 1.176 & 0.139 & 55 & 1.295 & 1.310  & \cite{PSRJ1829mass}\\
 J0453+1559  & 45.78 & 0.010 & 0.19 & 4.072  & 0.113 & $10^{3}$ & 1.559 & 1.174 &  \cite{Martinez15}\\
 J1411+2551 & 62.45 & 0.008 & 0.10 & 2.616 & 0.170 & 466 & \multicolumn{2}{c}{$m_{\text{tot}}=2.538 M_{\odot}$} & \cite{BNS1411}\\
  J1753$-$2240 & 95.14 & 0.005 & 0.97 & 13.64 & 0.304 & $10^{4}$ & -- & -- & \cite{Keith09} \\
 J1811$-$1736  & 104.2 & 0.004 & 0.90 & 18.78 & 0.828 & $10^{3}$ & \multicolumn{2}{c}{$m_{\text{tot}}=2.57 M_{\odot}$}  & \cite{Corongiu07}\\
 J1930$-$1852 & 185.5 & 0.003 & 18.0 & 45.06 & 0.399 & $10^{5}$ & \multicolumn{2}{c}{$m_{\text{tot}}=2.59 M_{\odot}$}  & \cite{Swiggum1930}\\
 J1755$-$2550 & 315.2 & 0.001 & 2430 & 9.696 & 0.089 & $10^{3}$ & -- & -- & \cite{Ng18psrJ1755}\\
  \hline
  \hline
 \end{tabular}
 \caption{Properties of 19 known Galactic BNS systems, including the spin period ($P$), dimensionless spin ($\chi$) and spin-down rate ($\dot{P}$) of the pulsar, binary orbital period ($P_b$), orbital eccentricity ($e_0$), merger time $T_{\rm{m}}$. Also listed are the mass of the recycled ($m_r$) and slow NS ($m_s$) in the binary, or the binary total mass ($m_{\text{tot}}$). The systems are ordered based on pulsar spin periods, with merging (non-merging) binaries listed in the top (bottom) half. For 12 entries with mass measurements for both stars, $m_r$ is the mass of the recycled pulsar, except PSR J1906+0746, which is a young pulsar (labeled as $m_s$ and its companion as $m_r$). For two globular-cluster pulsars (B2127+11C and J1807$-$2500B), their undetected companions could be either young or recycled NSs. In their respective references, PSRs J1906+0746, J1807$-$2500B, J1753$-$2240, and J1755$-$2550 were suggested to be likely in BNS systems, but their undetected companion stars could also be white dwarfs.}
  \label{tab:BNSspin}
\end{table}

\section{Comparison with LVC results under the low-spin prior}
\label{sec:appenB}

In this section, we compare our results to the public data release of the LVC under the $(m_{1},m_{2})$ parameterization.
(GW170817 data are available at \url{https://dcc.ligo.org/public/0150/P1800061/011/}, and GW190425 at \url{https://dcc.ligo.org/LIGO-P2000026/public}.)
Specifically, we compare to LVC results based on the low-spin prior, with component spin magnitudes $\chi_{1,2}$ uniformly distributed between 0 and 0.05.
The limit in this prior corresponds to the dimensionless spin of the fastest-spinning pulsar J1946+2052 \citep{Stovall18BNS} in \textit{known} merging Galactic BNS systems, whereas the high-spin prior ($\chi_{1,2}<0.89$) is an unconstrained prior with an upper end that is a technical limit imposed by available rapid waveform models \citep[e.g.,][]{gw170817}.

Our astrophysical spin prior within the $(m_{r},m_{s})$ paradigm improves on the simple LVC low-spin prior in two ways.
First, it incorporates the general expectation that the second-born NS is effectively nonspinning at merger for the standard BNS formation scenario.
We denote this NS as the slow NS and set $\chi_{s}=0$.
Second, for the first-born recycled NS, we make use of the empirical $P$-$P_{b}$ correlation, given in equation (\ref{eq:P0Pd}), which is derived by fitting pulsar measurements of Galactic BNSs within the framework of theoretical modeling of standard BNS formation \citep{Tauris15,Tauris17_BNS}.
By further assuming a log-uniform distribution for the initial BNS orbital period $P_b$, our model accounts for the likely distribution of dimensionless spins ($\chi_r$) of the recycled NSs for BNS mergers.
Our prior extends beyond the LVC low-spin limit, with $8\%$ of probability for $\chi_{r}>0.05$.
The low probability arises from the fact that the first-born NS is unlikely to get fully recycled, e.g., to achieve spin periods $\lesssim \unit[10]{ms}$, based on current understanding of BNS formation through isolated binary evolution.
We note that the cut on LVC low-spin prior at $\chi=0.05$ is set as a conservative limit in terms of the NS equation of state.
The current spin period of PSR J1946+2052 is $\unit[17]{ms}$, for which a dimensionless spin of 0.05 would correspond to an NS radius of $\sim \unit[15]{km}$.
(Note that the pulsar is spinning down slowly, likely to $\unit[18]{ms}$ when the BNS merges in $\sim \unit[46]{Myr}$.)
On the other hand, our prior on $\chi_r$ is, roughly speaking, marginalized over unknown NS radii over the plausible range of $[10,14]\,{\text{km}}$ \citep{Landry20NSradius}.

\begin{table}[!htb]
\centering
 \caption{Measured source parameters of GW170817 and GW190425 from this work, compared to those published by the LVC. We select parameters most relevant to this study, including the detector-frame binary chirp mass ($\mathcal{M}^{\text{det}}$), the source-frame chirp mass ($\mathcal{M}$), the source-frame binary total mass ($m_{\text{tot}}$), the mass ratio ($q$) and the effective spin parameter ($\chi_{\text{eff}}$). All quoted numbers are median posterior estimates and 5\% lower and 95\% upper limits, except $q$ for which the 10-100\% credible interval is given. Source-frame masses are obtained assuming the standard flat $\Lambda$CDM cosmology with a Hubble constant $H_0=\unit[67.9]{km\,s^{-1}\,Mpc^{-1}}$ and matter density parameter $\Omega_m=0.306$ \citep{Planck15cosmo}, except the LVC results for GW170817 where electromagnetic measurements of source redshift are used \citep{GW170817properties}. All results are based on the \texttt{IMRPhenomPv2\_NRTidal} waveform model. Additional details for such a comparison are shown in Figures (\ref{fig:mass_ratio}-\ref{fig:mrs_m12}). Full posterior samples are publicly available at \url{https://github.com/ZhuXJ1/BNSastro}.}
  \label{tab:LVCcompare}
 \begin{tabular}{l|c|ccccc}
  \hline
  \hline
  & & $\mathcal{M}^{\text{det}}$ & $\mathcal{M}$ & $m_{\text{tot}}$ & $q$ & $\chi_{\text{eff}}$  \\
  \hline
\multirow{2}{*}{GW170817} & this work  & $1.1975_{-0.0001}^{+0.0001}$ ${M_{\odot}}$ & $1.187_{-0.002}^{+0.004}$ ${M_{\odot}}$ & $2.73_{-0.01}^{+0.02}$ ${M_{\odot}}$ & $(0.79,1)$ & $0.00_{-0.01}^{+0.01}$   \\
& LVC  & $1.1975_{-0.0001}^{+0.0001}$ ${M_{\odot}}$ & $1.186_{-0.001}^{+0.001}$ ${M_{\odot}}$ & $2.73_{-0.01}^{+0.04}$ ${M_{\odot}}$ & $(0.73,1)$ & $0.00_{-0.01}^{+0.02}$  \\
\hline
\multirow{2}{*}{GW190425} & this work  & $1.4867_{-0.0003}^{+0.0003}$ ${M_{\odot}}$ & $1.44_{-0.02}^{+0.02}$ ${M_{\odot}}$ & $3.31_{-0.05}^{+0.05}$ ${M_{\odot}}$ & $(0.80,1)$ & $0.008_{-0.009}^{+0.015}$   \\
& LVC  & $1.4868_{-0.0003}^{+0.0003}$ ${M_{\odot}}$ & $1.44_{-0.02}^{+0.02}$ ${M_{\odot}}$ & $3.31_{-0.05}^{+0.06}$ ${M_{\odot}}$ & $(0.78,1)$ & $0.013_{-0.013}^{+0.014}$  \\
  \hline
  \hline
 \end{tabular}
\end{table}

\begin{figure*}[ht]
\begin{minipage}[t]{0.5\textwidth}
\includegraphics[width=0.9\linewidth]{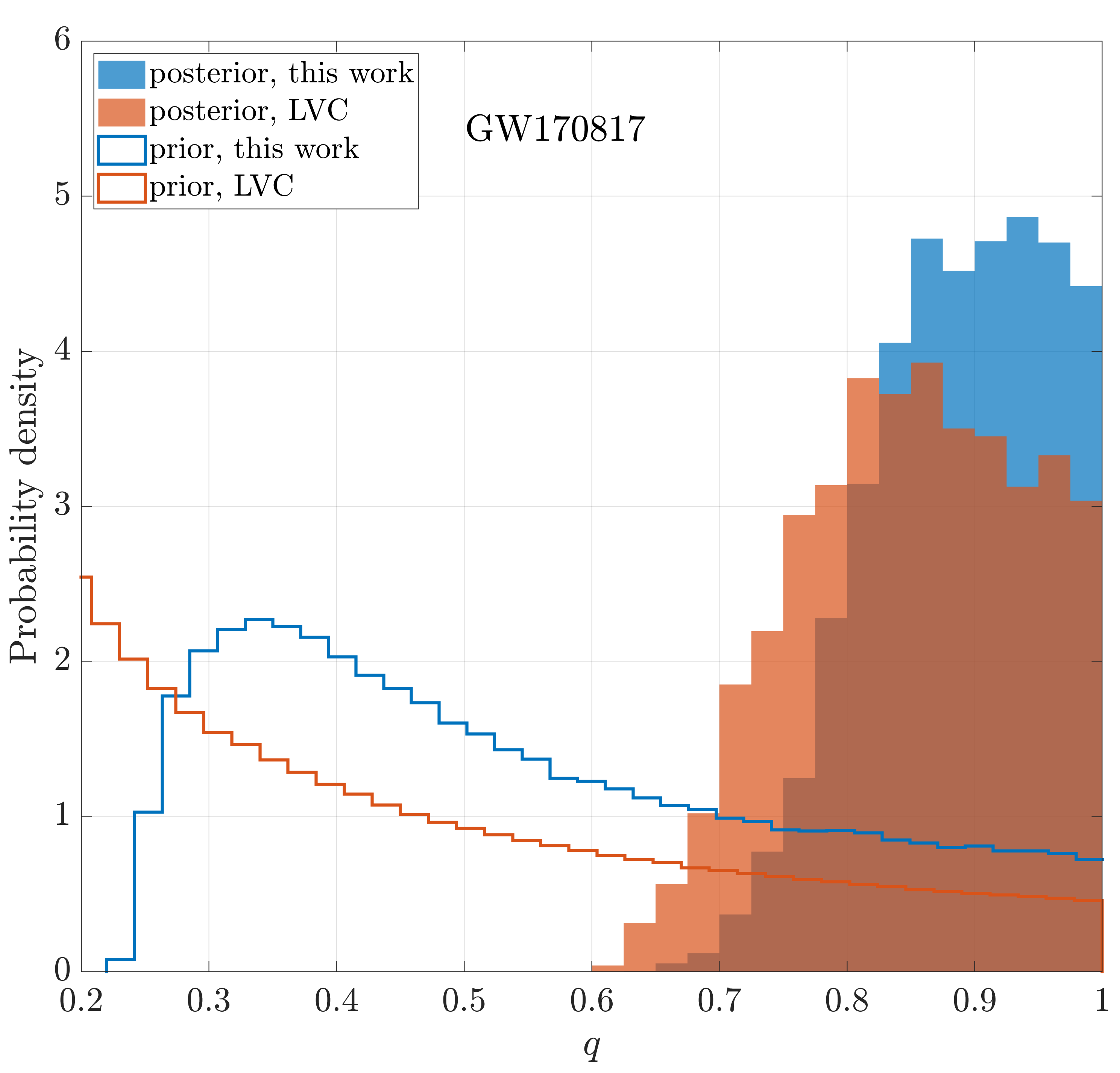}
\end{minipage}
\begin{minipage}[t]{0.5\textwidth}
\includegraphics[width=0.9\linewidth]{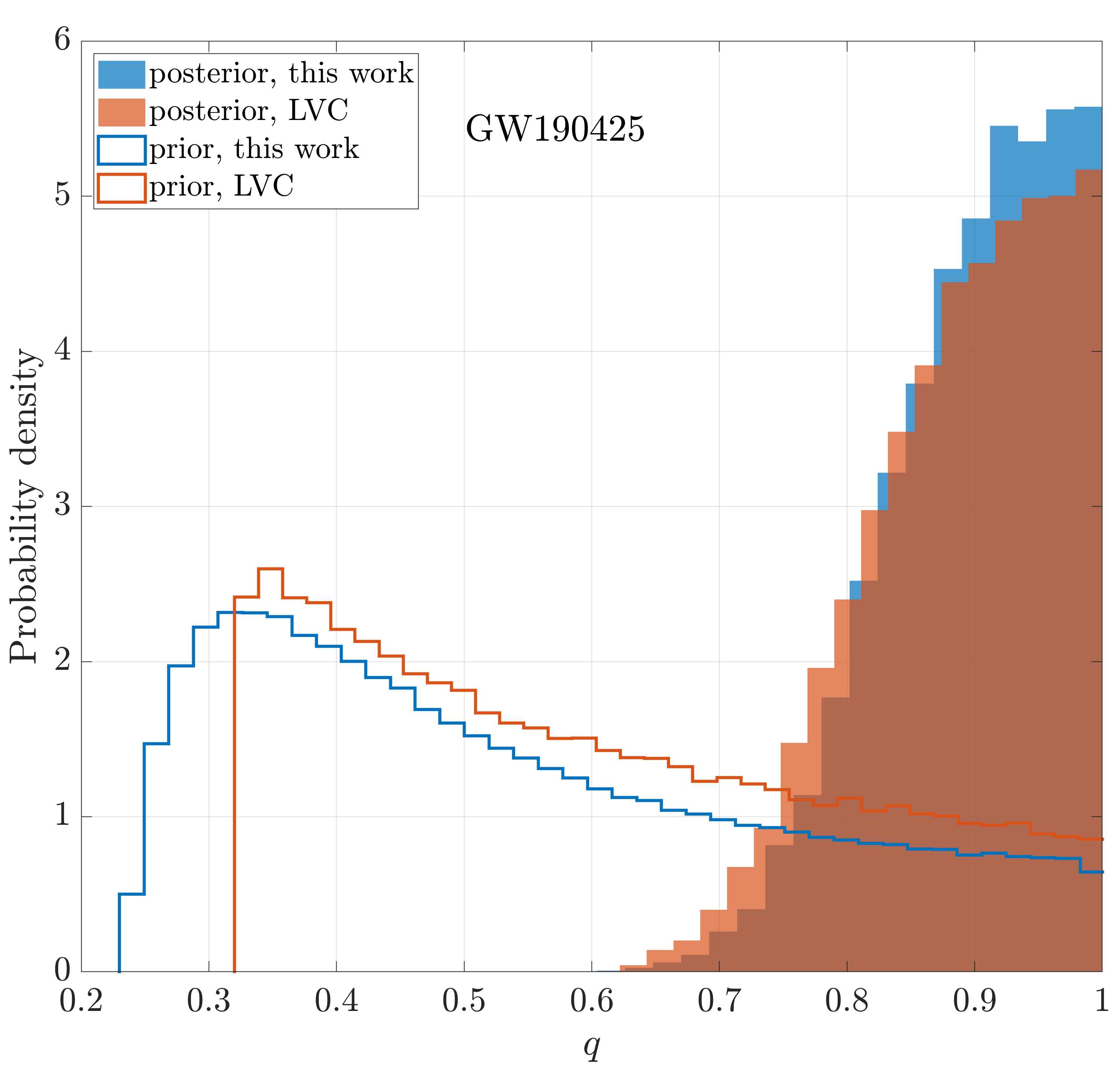}
\end{minipage}
\caption{Posterior probability distributions of the mass ratio ($q$) for GW170817 (\textit{left}) and GW190425 (\textit{right}), along with the prior distributions. We compare results from this work to the public data release by the LVC. The LVC mass priors are as follows. For GW170817, the detector-frame component masses are uniform in $[0.5,\, \unit[7.7]]{M_{\odot}}$, with $\mathcal{M}^{\text{det}}$ restricted in $[1.184,\, \unit[2.168]]{M_{\odot}}$ \citep{GW170817properties}. For GW190425, the detector-frame component masses are uniform in $[1.0,\, \unit[5.31]]{M_{\odot}}$, with $\mathcal{M}^{\text{det}}$ restricted in $[1.485,\, \unit[1.490]]{M_{\odot}}$ \citep{gw190425}. Both the LVC mass priors and ours are chosen for technical reasons. While the two sets of priors are different, we note that our effective priors on $q$ are nearly identical to those of the LVC for both events within the range spanned by the posterior distributions.}
\label{fig:mass_ratio}
\end{figure*}

\begin{figure*}[ht]
\begin{minipage}[t]{0.5\textwidth}
\includegraphics[width=0.9\linewidth]{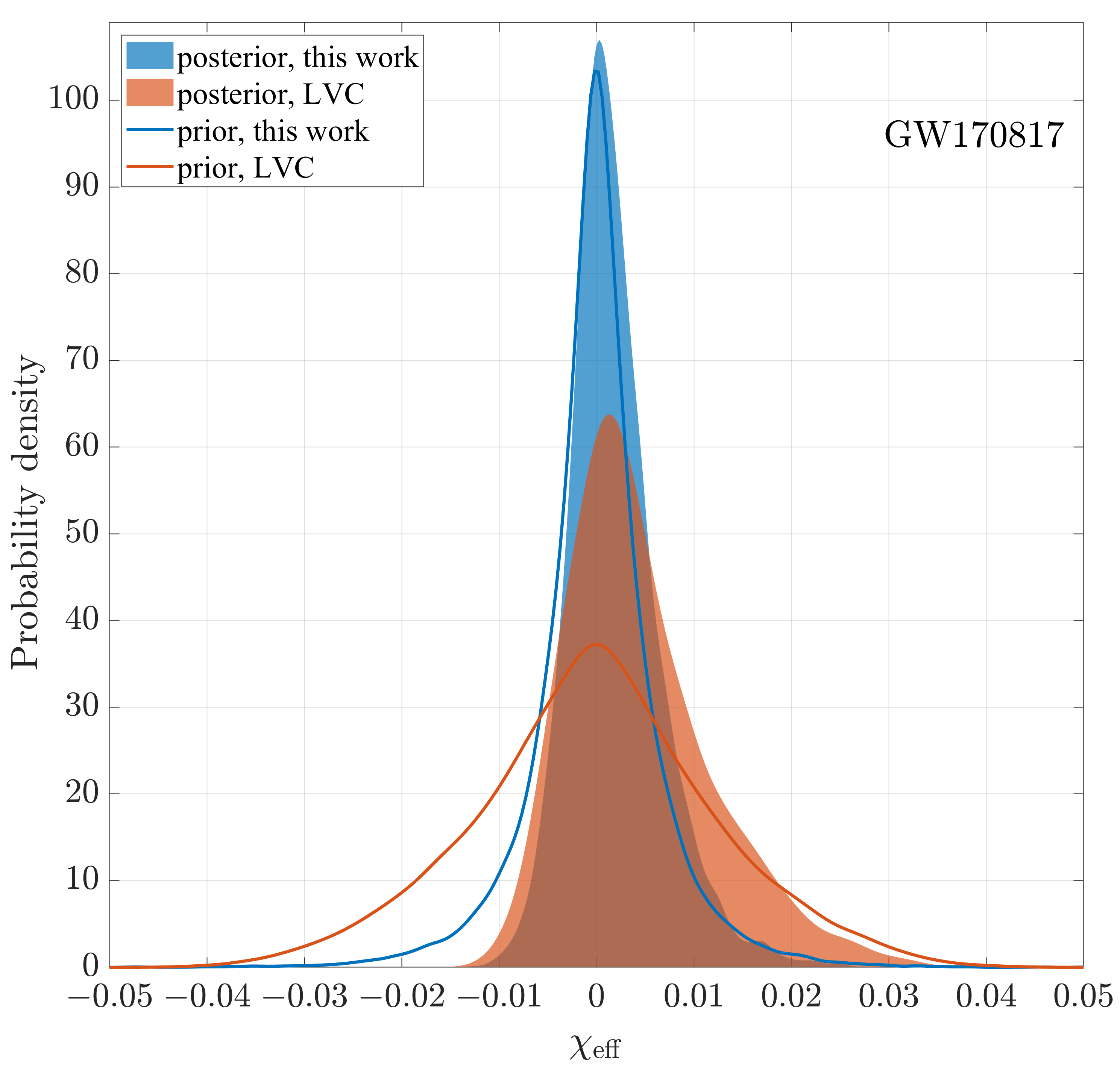}
\end{minipage}
\begin{minipage}[t]{0.5\textwidth}
\includegraphics[width=0.9\linewidth]{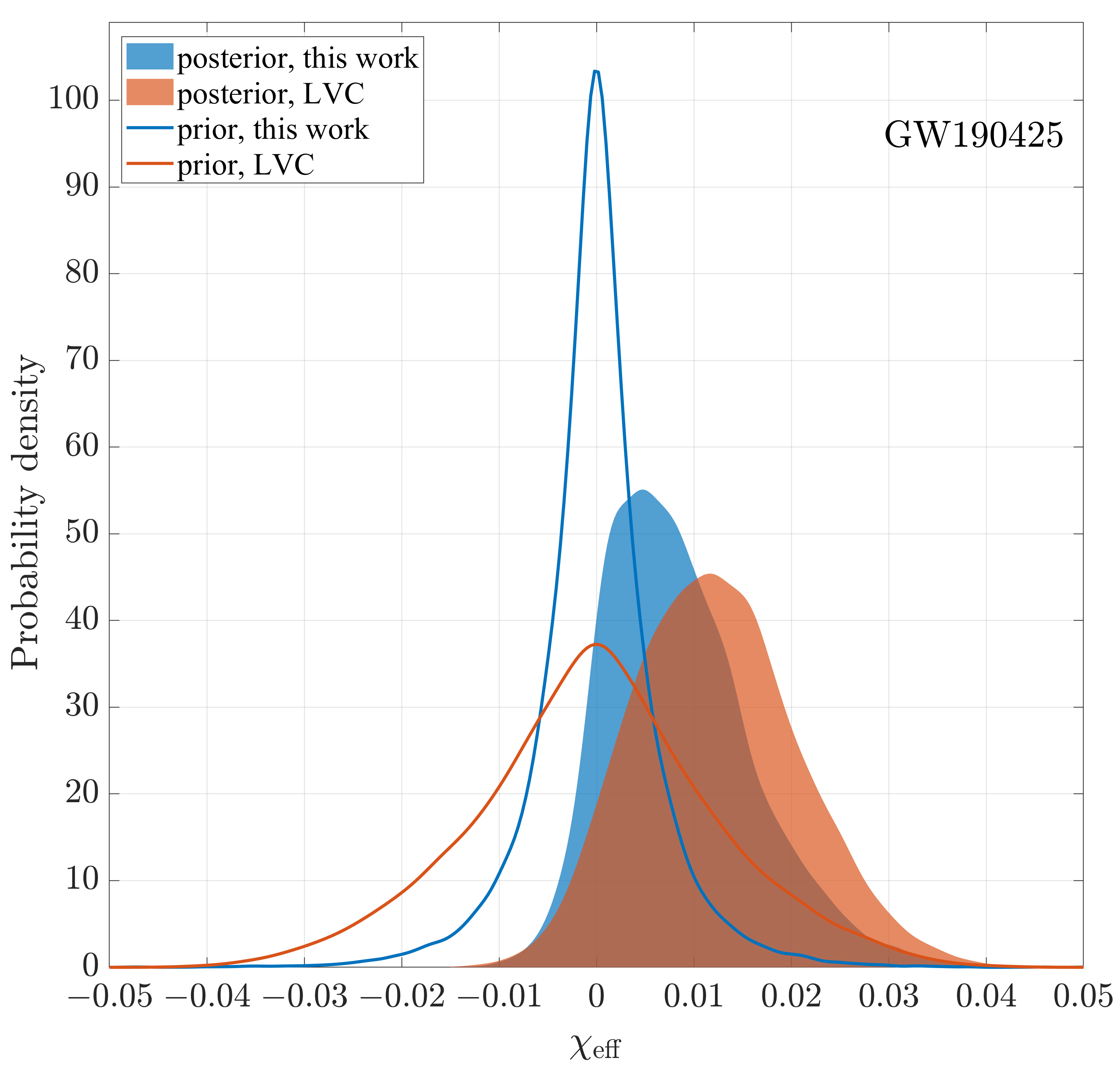}
\end{minipage}
\caption{Same as Figure \ref{fig:mass_ratio}, but for the effective spin parameter ($\chi_{\text{eff}}$).}
\label{fig:chi_eff}
\end{figure*}

In Table \ref{tab:LVCcompare}, we summarize posterior estimates of several common source parameters between this work and the LVC results.
Our measurements of detector-frame chirp mass $\mathcal{M}^{\text{det}}$, source-frame chirp mass $\mathcal{M}$, and total mass $m_{\text{tot}}$ are in excellent agreement with LVC results for both events.
The only exception is $\mathcal{M}$ of GW170817, due to different conversions from $\mathcal{M}^{\text{det}}$.
By incorporating electromagnetic information on the source luminosity distance of GW170817, the measurement uncertainty of $\mathcal{M}$ is reduced by a factor of 2 in \citet{GW170817properties}, from earlier analysis reported in \citet{gw170817}.
Our result agrees well with the latter in this case.

\begin{figure*}[ht]
\begin{minipage}[t]{0.5\textwidth}
\includegraphics[width=0.9\linewidth]{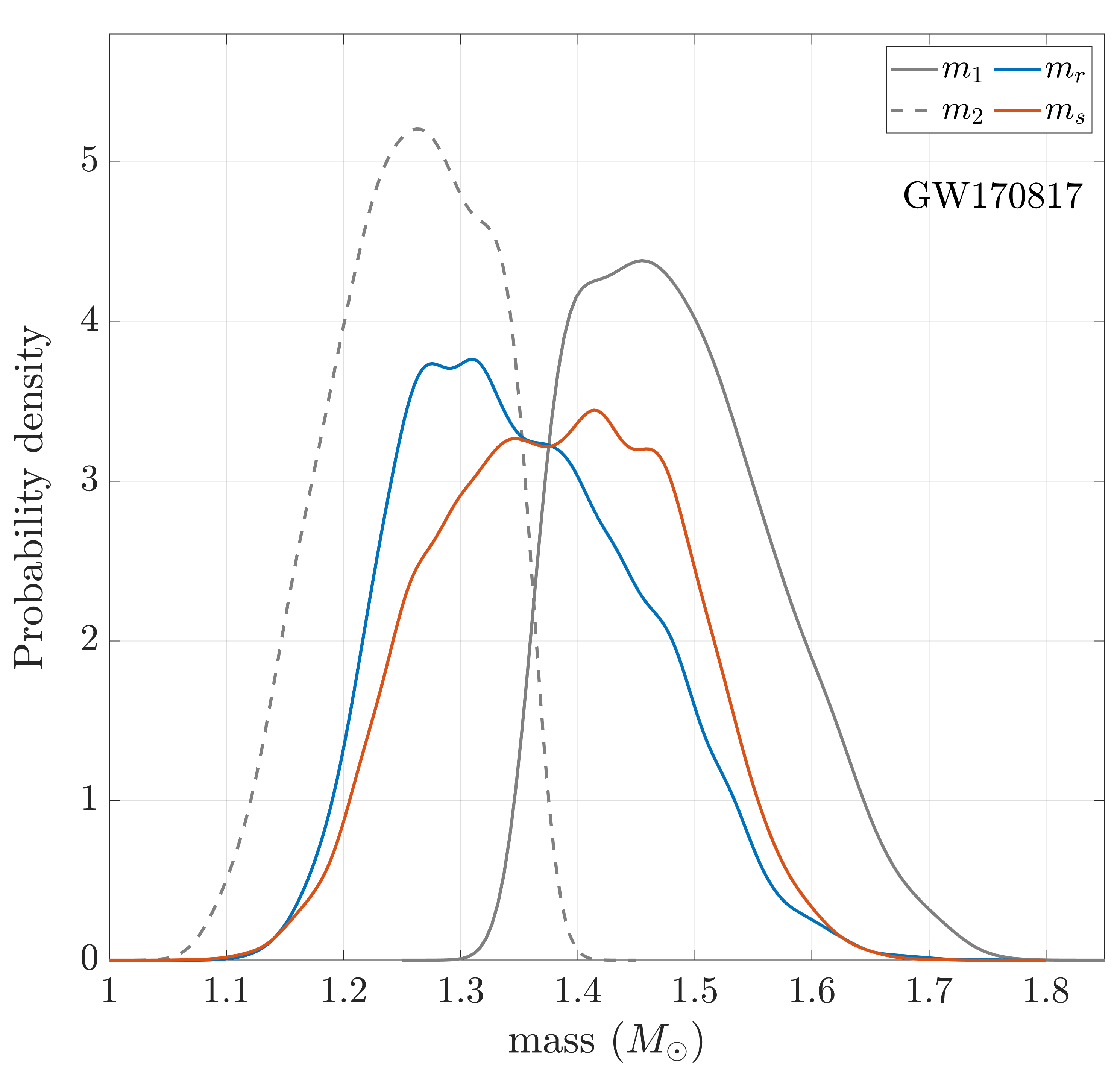}
\end{minipage}
\begin{minipage}[t]{0.5\textwidth}
\includegraphics[width=0.9\linewidth]{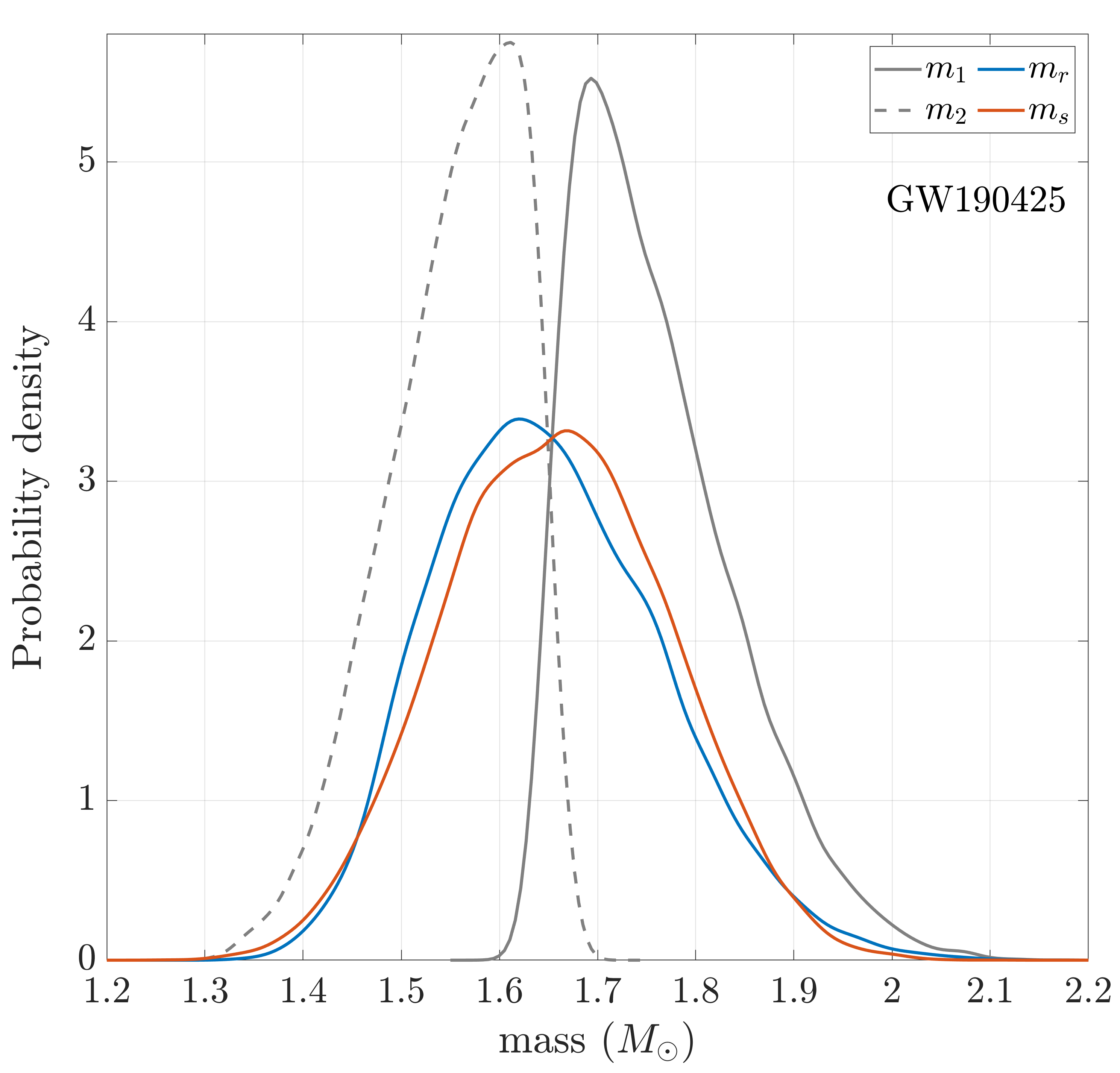}
\end{minipage}
\caption{Posterior probability distributions of source-frame component masses for GW170817 (\textit{left}) and GW190425 (\textit{right}): $(m_{r},m_{s})$ derived in this work, and $(m_{1},m_{2})$ published by the LVC. For GW170817, the LVC only released detector-frame component masses, which are converted to the source frame in the left plot using the redshift $z=0.009$ of its host galaxy NGC 4993 \citep{gw170817multi}.}
\label{fig:mrs_m12}
\end{figure*}

We present the posterior distributions along with priors for the mass ratio $q$ and effective spin parameter $\chi_{\text{eff}}$ in Figures \ref{fig:mass_ratio} and \ref{fig:chi_eff}, respectively.
We find our measurements of $q$ improve upon those of LVC, mildly for GW190425 but considerably for GW170817.
This can be understood as follows.
Our spin prior is more constraining, as can be seen in Figure \ref{fig:chi_eff}.
Because there is no support for measurable spins in GW170817 as found in Section \ref{sec:2events}, our prior constraint on $\chi_{\text{eff}}$ helps break the $\chi_{\text{eff}}-q$ degeneracy \citep[see, e.g., fig 7 in][]{GW170817properties}.
However, we find considerable support of spins for GW190425, but the data are not sufficiently informative to distinguish our prior from that of LVC due to the relatively low signal-to-noise ratio (12.9).
This leads to similar $\chi_{\text{eff}}$ posteriors and hence no significant difference in constraints on $q$.
Figure \ref{fig:mrs_m12} shows the component mass measurements for GW170817 (left) and GW190425 (right).

\bibliography{references}

\end{CJK}
\end{document}